\documentclass{article}

\newcommand{\beginsupplement}{%
    \setcounter{section}{0}
    \renewcommand{\thesection}{S\arabic{section}}%
    \setcounter{equation}{0}
    \renewcommand{\theequation}{S\arabic{equation}}%
    \setcounter{figure}{0}
    \renewcommand{\thefigure}{S\arabic{figure}}%
    \setcounter{table}{0}
    \renewcommand{\thetable}{S\arabic{table}}%
    \setcounter{page}{1}
    \renewcommand{\thepage}{S\arabic{page}}%
}

\usepackage{epsfig}
\usepackage{psfrag}
\usepackage{amssymb,amsmath}
\usepackage{bbold}
\usepackage{bm}
\usepackage{color}
\usepackage[T1]{fontenc}
\usepackage{empheq}
\usepackage{graphicx}
\usepackage{subcaption}
\usepackage{bm}
\usepackage{changes}

\newcommand{\be}{\begin{equation}}
\newcommand{\ee}{\end{equation}}

\def\wnem{w_{\rm n}}

\def\pt{{\partial}}
\def\ev{{\bf e}}
\def\fv{{\bf f}}

\def\nv{{\bf n}}

\def\wv{{\bf w}}

\def\pv{{\bf p}}

\def\tv{{\bf t}}

\def\tv{{\bf t}}
\def\uv{{\bf u}}
\def\vv{{\bf v}}

\def\gv{{\bf g}}

\def\Gv{{\bf G}}

\def\Lv{{\bf L}}

\def\Pv{{\bf P}}
\def\Qv{{\bf Q}}

\def\dd{{\rm d}}

\def\tr{{\rm tr}}

\def\grads{\nabla\hspace{-1mm}_s}
\def\Deltas{\Delta_s}

\def\dv{{\rm div}}

\def\dvs{\dv \hspace{-0.5mm}_s}
\def\esse{{\cal{S}}}

\def\astar{a^\star}

\newcommand{\ot}{\otimes}

\newcommand{\La}{\bm{\Lambda}}

\newcommand{\bnu}{\bm{\nu}}

\def\bzeta{{\bm{\zeta}}}

\def\QQ{{\mathcal{Q}_1}}

\newcommand{\sqrtone}{\sqrt{r^2+r_{,\vartheta}^2}}
\newcommand{\sqrttwo}{\sqrt{r^2\sin^2\vartheta+r_{,\varphi}^2}}
\newcommand{\sqrtthree}{\sqrt{\left(r^2+r_{,\vartheta}^2\right)\sin^2\vartheta+r_{,\varphi}^2}}
\newcommand{\one}{(r^2+r_{,\vartheta}^2)}
\newcommand{\three}{\left[(r^2+r_{,\vartheta}^2)\sin^2\vartheta+r_{,\varphi}^2\right]}
\newcommand{\Deltastar}{\Delta^{\hspace{-0.05mm}0}}
\newcommand{\gradstar}{\nabla^0_{\hspace{-1mm}s}}

\def\eeta{{ }}

\newcommand{\e}{\bm e}
\newcommand{\normal}{\bm{\nu}}
\newcommand{\n}{\bm n}

\begin{document}
\title{\bf Nematic Bubbles and the Breaking of Spherical Symmetry}
\author{
Gaetano Napoli\thanks{Dipartimento di Matematica e Applicazioni “Renato Caccioppoli”, Università degli Studi di Napoli “Federico II”, Napoli, Italy.} $\qquad$ Silvia Paparini\thanks{Dipartimento di Matematica “Tullio Levi-Civita”, Universit`a degli Studi di Padova, Padua, Italy.}
}
\date{}
\maketitle

\begin{abstract}
The emergence of nematic order on deformable closed surfaces plays a pivotal role in the morphogenesis of active biological matter, such as the regeneration of Hydra. In this work, we present a continuum model that couples the two-dimensional Landau-de Gennes order tensor, describing in-plane nematic ordering, with the mechanics of a mass-conserving, deformable spherical shell. By investigating the isotropic-to-nematic phase transition driven by a reduction in temperature—mimicking the natural induction of nematic order in actomyosin fibres—we perform both linear and weakly non-linear bifurcation analyses. The onset of nematic ordering spontaneously breaks spherical symmetry, yielding distinct equilibrium morphologies governed by the shell's deformability. Axisymmetric configurations, featuring two +1 defects at the poles, emerge via a discontinuous bifurcation, resulting in a globally stable prolate shape, alongside a metastable oblate shape. Non-axisymmetric configurations, featuring four +1/2 defects arranged in a square, arise via a continuous bifurcation. Shell softness drives the first-order character of the transition, while in the limit of infinite stiffness all bifurcations become continuous. Integer defects strongly couple with local mass redistribution—manifesting as shell thinning or thickening—whilst half-integer defects induce no such local deformation. These findings provide a purely mechanical framework for understanding body-axis formation and defect-mediated morphogenesis in biological vesicles.
\end{abstract}

\maketitle


\section{Introduction}
The study of crystalline order on curved surfaces, whether rigid or deformable, constitutes a compelling area of research, motivated both by its mathematical elegance and by its relevance to soft and biological matter. 
This field uniquely combines properties of geometry and topology to describe experimentally observed patterns and phenomenology, and to predict novel emergent effects. In the specific context of active biological matter, these fundamental principles often manifest through the emergence of liquid crystal phases.
Over the past two decades, a number of 2D realisations of biological systems \cite{keber:topology,kumar:tunable,saw:topological,blanch:turbulent,narayan:long} have shown that the long-range ordering of cells and cytoskeletal structures can result in a nematic phase, where elongated components align parallel to each other, inducing a partial orientational order akin to that observed in nematic liquid crystals.

Nematic liquid crystals consist of aggregates of rod-like molecules that exhibit orientational order while lacking positional order of their centres of mass. This orientational order is described by a unit vector $\nv \in \mathbb{S}^2$, known as the \emph{director}, which points along the mean molecular orientation. 
Whilst the classical framework adequately describes most nematic phenomena, the theory shows limitations in contexts directly relevant to this work. Indeed, a key feature of nematic systems is the presence of singularities in the director field, known as \emph{topological defects}, which can be classified by their topological charge \cite{virga:variational,ball:defects}. The director theory yields unsatisfactory predictions for the microscopic description of defects and surface effects, and it also fails to capture the physics of transitions between ordered and disordered phases, for example, when lowering the temperature of a mesogenic substance initially in the isotropic (disordered) phase. 
The order-tensor theory introduced by de Gennes \cite{degennes:physics} provides a more comprehensive description by focusing on the orientational probability distribution of the molecules and by introducing quantitative measures of the degree of orientational order and biaxiality. Within this framework, nematic liquid crystals admit three distinct phases, which may be identified through their optical properties, as the associated Fresnel ellipsoid is directly related to the order tensor itself. An isotropic liquid crystal is characterised by an isotropic order tensor and therefore behaves optically as an ordinary fluid. A uniaxial nematic phase possesses a single optic axis, corresponding to an order tensor with two coincident eigenvalues. By contrast, in a biaxial nematic phase all three eigenvalues of the order tensor are distinct, and the Fresnel ellipsoid exhibits two optic axes.

When nematic liquid crystals are constrained to a curved surface, the geometry induces a distortion in the molecular orientation. The transition to strict in-plane ordering depends on the shell thickness \cite{vitelli2006:nematic, lopezleon2011:frustrated}. In ultra-thin shells, the interaction with the colloid surface enforces a sort of degenerate anchoring, i.e., the tendency of the molecules to align along any direction parallel to the surface. Thus, unavoidable defects arise when nematic order is established on a surface with a spherical topology. The number of defects is a consequence of the Poincaré-Hopf theorem \cite{poincare1886:courbes,stoker1989differential,Hopf1983:differential}, which states that any configuration must have a total topological charge equal to the Euler-Poincaré characteristic of the surface. For instance, on a sphere it must be equal to $+2$.

Recent experiments on block copolymer vesicles with liquid-crystalline side chains reveal that the delicate interplay between the two-dimensional ordering, which exhibits topologically required defects, and vesicle shape can lead to a rich variety of vesicle morphologies \cite{xu2009:selfassembly,boisse2009:synthesis,delbarrio2010:selfassembly,yang2010:amphiphilic}. Here, defects play a central role in determining the resulting vesicle morphology. Moreover, it has been observed in developing organisms that topologically required defects in the nematic order act as organising centres to grow protrusions or deplete material to relieve stress \cite{copenhagen:topological,kawaguchi:topological,keber:topology,maroudas2020:topological,saw:topological}.
An example is \emph{Hydra}, a biologically immortal organism in which topological defects align with morphological features: a defect of charge $+1$ is localised at the tip of each tentacle, while two $-1/2$ defects reside at its base \cite{maroudas2020:topological}.

Over the past two decades, various models have been proposed to understand the unique environment surrounding a defect that drives the fundamental morphological developments \cite{wang:patterning,duclut:active,kramer:biological,Khoromskaia2023:active,paparini2025:shapeinstabilities,carenza2022:theory,metselaar2019:topology,vafa2023:active,Maroudas2025:mechanical,hernandez2026:mechanics}. Modelling such systems remains a daunting task due to the numerous interacting mechanisms involved, including two-phase flow, active stresses localised at the interface, and the coupling between biochemical signalling, shape, and curvature that drives pattern formation. 

In this work, we are primarily interested in the early stages following the excision of a fragment of \emph{Hydra}. In \cite{maroudas2020:topological}, the nematic dynamics of actin fibres during \emph{Hydra} regeneration are monitored by live microscopy. It is observed that, initially, an excised tissue fragment from a mature \emph{Hydra} exhibits a nearly perfect nematic arrangement of actin fibres inherited from the parent organism.
The excised tissue subsequently folds, within a few hours, through an active actomyosin-dependent process, forming a closed spheroid in which the nematic organisation of the fibres can be assumed to be essentially lost. 
Subsequently, nematic arrays of actin fibres gradually form 
and topological constraints necessarily give rise to defects in their alignment. 
This induction of order into the fibres confers a structural memory of the original body axis orientation, in accordance to which future head and foot form.
The details of how body-axis polarity memory is encoded in the tissue remain unknown, but it is apparent already at the earliest stages of the regeneration process. Its underlying physical mechanisms based on the interplay between biochemical and mechanical activity are the subject of theoretical investigation. In \cite{wang:patterning}, axis formation is driven by the alignment of nematic fibres with a global morphogen concentration field, by analogy with nematic liquid crystals under external fields. To overcome the conflicts arising with observations that Wnt (family of signalling proteins) is localised mainly in the head region, in \cite{Maroudas2025:mechanical} a strain-regulated morphogen expression is proposed: higher strain increases morphogen concentration, reinforcing fibre alignment along its gradient. Instead, in \cite{hernandez2026:mechanics}, the regenerating \emph{Hydra} is modelled as an active spherical shell, where the interplay of active nematic stresses and shell elasticity spontaneously identifies a body axis, aligning muscle fibres without a pre-existing morphogen gradient.

Our work is set within this framework to investigate the molecular origin of muscle fibre alignment and its interplay with the mechanics of the deformable surface during \emph{Hydra} regeneration. We examine the effect of cooling a nematic confined within a deformable spherical shell, initially isotropic, representing the spheroid formed after the excised \emph{Hydra} folds onto itself. Cooling mimics the natural induction of nematic order in muscle fibres, inducing a nematic phase in the shell. We then analyse the onset of nematic ordering, the resulting surface deformation, and how this symmetry-breaking transition may lead to head-foot formation.

More precisely, this paper introduces a continuum model of a deformable surface coated with a nematic liquid crystal exhibiting degenerate planar anchoring, featuring a strong coupling between surface deformation and the tangential nematic texture. While the mechanics of the surface are primarily governed by surface tension $\gamma$ and sustained by an internal pressure $\mu$—analogous to a soap bubble—the presence of the nematic order fundamentally alters the system. This occurs through a twofold mechanism: energetically, the confinement of liquid crystal molecules to the tangent plane imparts an effective bending elasticity; topologically, the spherical genus necessitates the formation of defects, which inevitably break the perfect spherical symmetry and induce shape distortion.

The in-plane nematic ordering is described by using the two-dimensional Landau–de Gennes order tensor introduced in \cite{Biscari:2006, kralj:2011}:
\begin{equation}
\Qv_s = q \left(\nv \ot \nv - \frac{\Pv}{2}\right),
\label{Qtensor}
\end{equation}
where $\nv$ is a unit tangential vector field representing the nematic director, $q \in [-1,1]$ is the scalar order parameter, and $\Pv$ denotes the surface identity tensor, which acts as the projector onto the tangent plane.

In this study, we adopt a variational approach to derive the general non-linear equilibrium equations coupling nematic order to the underlying surface. These equations admit as a solution a spherical shape wherein the nematic is everywhere in the isotropic phase ($\Qv_s=\bf 0$). Given the pivotal role played by topological defects in shaping natural morphologies, this work investigates the effects of lowering the temperature of a nematic confined within a deformable spherical shell. We subsequently examine the isotropic–nematic transition as a temperature-induced bifurcation, accompanied by a loss of sphericity in the solutions. This analysis leads to the determination of quasi-spherical equilibrium shapes, their associated nematic textures, and the resulting defect configurations.

The paper is organised as follows. In Section~\ref{sec:energy_functional}, we introduce the energy functional $W$, which couples the classical surface energy of a soap bubble with the nematic distortion energy, modelled via the two-dimensional Landau-de Gennes framework. We subsequently derive the governing equilibrium equations as the Euler–Lagrange variations of $W$ defined on a closed manifold. In Section~\ref{sec:linearization}, we define our class of non-axisymmetric configurations, specifically quasi-spherical surfaces endowed with tangential nematic order. To investigate the bifurcation problem, we seek solutions to the equilibrium equations that are slightly perturbed with respect to the isotropic spherical configuration, where the perturbation is governed by a small parameter $\varepsilon \ll 1$. This formulation yields an eigenvalue problem, wherein the lowest eigenvalue determines the critical instability threshold and the corresponding bifurcation modes (eigenfunctions), expressed in terms of spherical harmonics and spin-weighted spherical harmonics. The critical eigenvalue is found to be degenerate, admitting both axisymmetric modes—characterised by two $+1$ defects positioned at the poles—and non-axisymmetric modes featuring four coplanar $+1/2$ defects. To determine the amplitude of the bifurcated solutions, we perform a weakly nonlinear analysis in Section~\ref{sec:weak_nonlinear} using a higher-order approximation of the free energy. This analysis reveals that the axisymmetric modes bifurcate via a transcritical transition, whereas the non-axisymmetric modes follow a supercritical transition. Finally, in Section~\ref{sec:conclusions}, we synthesise and discuss our main results, whilst technical details of the proofs are provided in the Electronic Supplementary Material (ESM).

\section{Model}\label{sec:energy_functional}

We model a nematic bubble as a regular closed surface $\esse$, topologically equivalent to a sphere, equipped at each point $\bm p$ with an outward unit normal $\bnu$. The underlying concept extends the classical model of the bubble---wherein the energy derived from surface tension $\gamma>0$ is minimised subject to a fixed enclosed volume $V_0$---by augmenting it with an additional term arising from the nematic phase. We therefore write:
\be
\label{eq:free_energy}
W[\esse] = \gamma\; {\rm area(\esse)} -  \mu \;\left( {\rm vol(\esse)} - V_0\right) + \int_{\esse}\rho \; \wnem\, \dd A,
\ee
where ${\rm area(\esse)}$ represents the total area of the surface $\esse$,  ${\rm vol(\esse)}$ denotes the volume enclosed by it, and $\mu$ is the Lagrange multiplier associated with the volume constraint. Furthermore, $\rho$ denotes the mass density (i.e., the mass per unit area), and $\wnem$ represents the nematic distortion energy density per unit mass. We detail these three contributions below. 

Using the divergence theorem, we may obtain the volume via an integral over the enclosing surface:
\be
{\rm vol(\esse)}=\frac{1}{3}\int_S \bm p\cdot \bnu \; \dd A,
\ee
thereby allowing us to express $W[\esse]$, up to an additive constant, purely as an integral over $\esse$:
\be
W[\esse] = \int_\esse \left(\gamma - \frac{\mu} {3} \bm p\cdot \normal + \rho \; \wnem\right)\, \dd A + \mu V_0.
\label{eq:W}
\ee
The nematic energy term is defined via an energy density per unit mass, rather than per unit area. This choice is consistent with the fact that the magnitude of the nematic contribution scales with the film thickness, which inevitably varies as the bubble changes size. Indeed, in the model presented in \cite{Napoli:2012}, the parameters scale with the film thickness. This variation in thickness is accounted for by $\rho$, which represents the volumetric mass density integrated across the thickness, effectively yielding the mass per unit area. Since the film constituting the bubble is three-dimensionally incompressible, its effective two-dimensional counterpart must be extensible to accommodate changes in bubble size, whilst strictly conserving mass.

\subsection{Nematic distortion energy}
The nematic order on the bubble surface is described by the two-dimensional order tensor \eqref{Qtensor}. Note that the value $q=0$ corresponds to local isotropic states within the tangent plane. When $q>0$, $\nv$ represents the nematic director; conversely, if $q<0$, the director corresponds to the conormal vector $\tv=\bnu \times \nv$. In other words, the order tensor associated with a negative value of $q$ and director $\nv$ is equivalent to that associated with a positive degree of order $-q$ and director $\tv$.

For the specific choice of $w_{\rm nem}$, we refer the reader to Ref.~\cite{Napoli:2012}. As is customary, we decompose this term into two contributions:
\be
\wnem=w_{\rm el}+w_{\rm LdG}.
\label{eq:nem}
\ee
The elastic term $w_{\rm el}$ is derived from the three-dimensional theory under the one-constant approximation. In the reduced 2D model, this formulation yields three contributions:
\be
w_{\rm el} = \frac{k}{2}\left(|\grads \Qv|^2 + 2 \eeta \Qv \cdot \Lv^2 + \frac{1}{2}|\Lv|^2\right),
\label{eq:wel}
\ee
where $k$ is a positive constant and $\Lv = -\grads \bnu$ denotes the extrinsic curvature tensor. In equation \eqref{eq:wel}, the first term associates an energetic cost with non-uniform states of $\Qv$, whilst the second couples the nematic director to the extrinsic curvature tensor $\Lv$, thereby favouring the alignment of the eigenvectors of $\Qv$ (and consequently the nematic director) with the principal directions of surface curvature. The third term represents a form of curvature elasticity imparted to the surface by the nematic field, arising from the constraint that the field must remain tangent to the surface. Notably, this contribution does not vanish even when the nematic is in the isotropic phase, as the two-dimensional isotropic phase on a curved surface is not represented by a spatially constant tensor.

Given that the quantity $|\grads \Qv|^2$ can be expressed in terms of the surface gradients of $\nv$ and $q$ as follows:
\be
|\grads \Qv|^2 = 2 q^2 |\grads \nv|^2 + \frac{1}{2}|\grads q|^2 - 2 q^2 |\Lv \nv|^2 + \frac{1}{2} q^2 |\Lv|^2,
\ee
we may rewrite $w_{\rm el}$ in the form:
\be
\label{eq:w_nem}
w_{\rm el} = \frac{k}{2} \Big[\underbrace{\frac{1}{2}|\grads q|^2  + 2 q^2 \left(|\grads \nv|^2 -  |\Lv \nv|^2 + \frac{1}{4}  |\Lv|^2\right)}_{|\grads \Qv|^2} + 2 \eeta \underbrace{q\left( |\Lv \nv|^2 - \frac{1}{2} \tr(\Lv^2)\right)}_{\Qv\cdot \Lv^2} + \frac{1}{2}|\Lv|^2 \Big].
\ee

The Landau–de Gennes energy density $w_{\rm LdG}$, which is typically thermotropic in nature, exhibits a quartic dependence on $q$ and is given by:
\be
\label{eq:w_LdG}
w_{\rm LdG} = \frac{a}{2} q^2 + \frac{c}{4} q^4,
\ee
where $a = a_0 (T - T_{\rm NI})/T_{\rm NI}$, with $a_0$ and $c$ being positive constitutive constants, and $T$ denoting the absolute temperature. The nematic-isotropic transition temperature $T_{\rm NI}$ is defined as the temperature at which, upon heating, a planar nematic undergoes a phase transition from the ordered nematic liquid crystal phase to the disordered isotropic phase.

\subsection{Derivation of equilibrium equations}
We now derive the equilibrium equations as extremals of the energy functional \eqref{eq:W}, by requiring that its first variation vanishes. Assuming that each point on the surface undergoes a normal virtual displacement defined by $\uv = \bnu \delta u $, we consequently obtain (see, for example, Ref.~\cite{napoli:2010}):
\begin{equation}
\grads \uv = -\delta{u} \; \Lv + \bnu \otimes \grads \delta{u}, \qquad
\grads^{\;\top} \uv = -\delta{u} \; \Lv + \grads \delta{u} \otimes \bnu, \qquad
\dvs \uv  = -2H\;\delta{u}.
\label{eq:varu}
\end{equation}
Furthermore, the surface unit normal and the extrinsic curvature tensor also undergo variations; these may be expressed as functions of $\delta u$ and its surface gradient as follows:
\be
\delta \bnu = - \grads \delta u, \qquad \delta \Lv = \Lv ( \grads \delta{u} \otimes \bnu ) + \grads ( \grads \delta{u} ) + \delta{u} \; \Lv^2.
\ee
The nematic director $\nv$ varies not only due to the deformation of the surface itself but also because $\nv$ may rotate by an angle $\delta \theta$ about the normal. Employing Eqs.~(16) and (17) of \cite{napoli:2010}, in conjunction with \eqref{eq:varu}, the variation of $\nv$ can be expressed as
\be
\delta \nv = (\nv \cdot \grads \delta{u}) \bnu + \tv \, \delta \theta,
\ee
and, consequently, the variation of its gradient yields:
\be
\delta ( \grads \nv ) = ( \grads \nv ) [ \grads \delta{u} \otimes \bnu - \bnu \otimes \grads \delta{u} ] + \delta{u} ( \grads \nv ) \Lv + \grads [ ( \nv \cdot \grads \delta{u} ) \bnu ] + \grads ( \tv , \delta \theta ).
\ee
Finally, let $\delta q$ denote the variation of the scalar order parameter $q$; invoking Eq.~(18) of \cite{napoli:2010} along with \eqref{eq:varu}, the variation of its gradient is given by:
\begin{equation}
\delta \grads q = ( \grads q \cdot \grads \delta{u} ) \bnu + \delta{u} \Lv \grads q + \grads \delta q.
\end{equation}

In computing the variation $\delta W$, we impose the condition of mass conservation:
\begin{equation}
\delta \int_{\esse} \rho \, \dd A = 0,
\label{eq:mc}
\end{equation}
yielding 
\begin{equation}
\int_{\esse} ( \delta \rho + \rho \dvs \uv ) \, \dd A = 0.
\label{eq:mc1}
\end{equation}
Combining the final expression of Eqs.~\eqref{eq:varu} with Eq.~\eqref{eq:mc1}, and invoking the arbitrariness of the surface $\esse$, we derive the local form of the mass conservation constraint:
\begin{equation}
\delta \rho - 2 \rho H \delta{u} = 0.
\label{eq:mc2}
\end{equation}

We now proceed to compute the variation of the functional defined in Eq.~\eqref{eq:W}:
\begin{equation}
\delta W = \int_{\esse} \left[ - 2 \gamma H \delta{u}-\frac{\mu}{3}\left(\delta\left(\bm p\cdot\normal\right)+\left(\bm p\cdot\normal\right)\dvs\bm u\right) + \rho \, \delta \wnem\right] \, \dd A.
\label{eq:deltaW}
\end{equation}
The variation $\delta \wnem$ follows from the chain rule:
\begin{equation}
\delta \wnem = \frac{\partial \wnem}{\partial \Lv} \cdot \delta \Lv + \frac{\partial \wnem}{\partial \nv} \cdot \delta \nv + \frac{\partial \wnem}{\partial \grads \nv} \cdot \delta \grads \nv + \frac{\partial \wnem}{\partial q} \, \delta q + \frac{\partial \wnem}{\partial \grads q} \cdot \delta \grads q, \nonumber
\end{equation}
which we shall compute term by term below. 
To this end, it is convenient to introduce the quantities:
\begin{equation}
\label{eq:flussi}
\La := \rho \frac{\partial \wnem}{\partial \Lv}, \quad \gv := \rho \frac{\partial \wnem}{\partial \nv}, \quad\Gv := \rho \frac{\partial \wnem}{\partial \grads \nv}, \quad \xi := \rho \frac{\partial \wnem}{\partial q}, \quad\bzeta := \rho \frac{\partial \wnem}{\partial \grads q}.
\end{equation}

Consequently, the contribution to $\delta W$ arising from the variation of $\Lv$ is
\begin{equation}
\int_{\esse} \La \cdot \delta \Lv \, \dd A = \int_{\esse} \left[ \dvs ( \dvs \La ) + \La \cdot \Lv^2 + 2H \bnu \cdot ( \dvs \La ) \right] \delta{u} \, \dd A,
\label{i1}
\end{equation}
where the result follows from successive applications of integration by parts and the surface divergence theorem:
\begin{equation}
\int_{\esse} \dvs \vv  \dd A = - 2\int_{\esse} H \vv \cdot \bnu \; \dd A,
\end{equation}
for any vector $\vv$ and closed surface $\esse$.

The contributions involving the variations $\delta \nv$ and $\grads(\delta \nv)$ can now be evaluated. Since $\gv$ is tangential, we get
\begin{equation}
\label{i2}
\int_{\esse} \gv \cdot \delta \nv \, \dd A = \int_{\esse} \gv \cdot \tv \, \delta \theta \, \dd A,\end{equation}
and
\begin{equation}
\label{i3}
\int_{\esse} \Gv \cdot \delta ( \grads \nv ) \, \dd A = \int_{\esse} \left\{ \dvs [ ( \bnu \cdot \dvs \Gv ) \nv ] + \Gv \cdot [ ( \grads \nv ) \Lv ] \right\} \delta{u} \, \dd A - \int_{\esse} \tv \cdot \dvs \Gv \, \delta \theta \, \dd A.
\end{equation}
Finally, the contribution of the variations of $q$ and $\grads q$ to $\delta W$ is
\begin{equation}
\label{i4}
\int_{\esse} ( \xi \, \delta q + \bzeta \cdot \delta \grads q ) \, \dd A = \int_{\esse} \bzeta \cdot ( \Lv \grads q ) \delta{u} \, \dd A + \int_{\esse} ( \xi - \dvs \bzeta ) \, \delta q \, \dd A,
\end{equation}
where it is understood that $\bzeta$ is tangential to $\esse$.

Upon substituting equations \eqref{i1}, \eqref{i2}, \eqref{i3}, and \eqref{i4} into \eqref{eq:deltaW}, and requiring that $\delta W = 0$ for arbitrary variations $\delta{u}$, $\delta \theta$, and $\delta q$, we obtain the following set of equilibrium equations:
\begin{subequations}
\label{eq:set_equations}
\begin{align}
{ -2\gamma H}+\dvs ( \Pv \dvs \La + \bnu \cdot \dvs \Gv \, \nv ) + \La \cdot \Lv^2 + \Gv \cdot ( \grads \nv ) \Lv + \bzeta \cdot \Lv \grads q { -\mu}&= 0\label{eq:shape}, \\
\tv \cdot ( \gv - \dvs \Gv ) &= 0,\label{eq:n2} \\
\xi - \dvs \bzeta &= 0\label{eq:q2}, 
\end{align}
\end{subequations}
where we used the identity
\begin{equation}
 \dvs ( \Pv \dvs \La) +  2 H \bnu \cdot \dvs \La =  \dvs ( \Pv \dvs \La).
\end{equation}

Equations \eqref{eq:set_equations}, together with the constraint
\be
\text{vol}(\esse) = V_0,
\label{eq:constraint}
\ee
constitute a system of scalar equations, where the unknowns are the surface shape, the nematic director, the scalar order parameter, and the Lagrange multiplier $\mu$. This system of equations is closed by the following constitutive relations:
\begin{subequations}
\begin{gather}
\La   =\frac{k\rho}{2} \left[ 2 q (1\eeta-q)(\Lv \nv \ot \nv + \nv \ot \Lv \nv)  + (q^2 - 2 \eeta q + 1)  \Lv \right]\\
\Gv =   2 \rho k q^2 \grads \nv, \qquad
\gv  = 2\rho k q  (1\eeta-q) \Lv^2 \nv, \qquad
\bzeta = \frac{k\rho}{2} \grads q,\label{eq:b}\\
\xi  = \rho\left[a q + c q^3 + \frac{k}{2}(2 \eeta |\Lv \nv|^2 - \eeta\Pv \cdot \Lv^2  + q(4 |\grads \nv|^2 - 4 |\Lv \nv|^2 +  |\Lv|^2))\right]\label{eq:p},
\end{gather}
\end{subequations} 
which follow from the definitions \eqref{eq:flussi} and the specific form of the nematic energy given by equations \eqref{eq:nem}, \eqref{eq:w_nem}, and \eqref{eq:w_LdG}.

\subsection{Ground state solution}\label{sec:ground_state}
We begin by considering a configuration in which the nematic liquid crystal is in the isotropic phase within the tangent plane; that is, $q = 0$ everywhere. Thus, from equation \eqref{eq:q2}, and utilising equation \eqref{eq:p},
we obtain
\be
H\varsigma=0,
\label{asph}
\ee
where $\varsigma = \nv \cdot \Lv \nv - \tv \cdot \Lv \tv$ is termed the \emph{asphericity}. Consequently, equation \eqref{asph} is satisfied globally on the surface in two instances: when $H=0$ (minimal surfaces), or when $\varsigma = 0$, which corresponds to a spherical geometry.

Assuming a spherical solution, equation \eqref{eq:constraint} determines its radius
\be
r_0 = \sqrt{\frac{3 V_0}{4 \pi}}.
\ee

The director equation \eqref{eq:n2} is then identically satisfied, whereas \eqref{eq:shape} reduces to
\be
\label{eq:shape_q0}
2H\gamma + \normal \cdot \dvs\left(\frac{k\rho}{2} \Lv^2\right) + \mu = 0.
\ee
For a sphere of radius $r_0$, the curvature tensor is given by $\Lv = {\rm diag}[-1/r_0, -1/r_0, 0]$ in any local basis, or equivalently $\Lv= H \Pv$, given that the mean curvature is $H = -1/r_0$. Moreover, assuming a constant surface mass density $\rho = \rho_0$, and in view of the identity $\dvs \Pv = -2H\normal$, equation \eqref{eq:shape_q0} yields the Lagrange multiplier $\mu$ via a modified \emph{Young-Laplace law}:
\be
\label{eq:YL}
\mu = \frac{2}{r_0}\left(\gamma - \frac{k\rho_0}{2r_0^2}\right).
\ee
The additional term, compared with the classical Young-Laplace law, arises from curvature elasticity. As previously noted, this stems from the fact that the tangential isotropic order is not represented by a constant tensor. The negative sign indicates that this elastic term acts in aid of the internal pressure. Since the curvature energy decreases with increasing radius, the bubble tends to expand spontaneously to achieve a flatter configuration. Consequently, a lower internal pressure $\mu$ is required to sustain the bubble than is predicted by the classical law.

\section{Linearisation about the ground state}\label{sec:linearization}
In this section, we derive the linearised equations about the ground-state configuration. These equations govern the equilibrium of nearly spherical shapes with a small scalar order parameter.

Let $\pv = x \ev_x + y \ev_y + z \ev_z$ be a point on the surface $\esse$, and consider the following parametrization in spherical coordinates:
\be
\label{eq:shell_par}
x = r(\vartheta,\varphi) \sin \vartheta \cos \varphi, \qquad
y = r(\vartheta,\varphi) \sin \vartheta \sin \varphi, \qquad
z = r(\vartheta,\varphi) \cos \vartheta,
\ee
where $\vartheta\in[0,\pi)$ and $\varphi\in[0,2\pi)$. Hence, a point $\pv$ on a surface $\esse$ in this class can be expressed in terms of the spherical frame $(\e_r, \e_\vartheta, \e_\varphi)$ as
\be
\label{eq:pointp}
\pv = r(\vartheta,\varphi) \, \ev_r.
\ee
This parametrisation allows for non-axisymmetric bubble shapes through the dependence of $r$ on both angular variables $\vartheta$ and $\varphi$. When $r$ is a constant $r_0$, the surface $\esse$ coincides with a sphere of radius $r_0$. If, instead, $r=r(\vartheta)$, the resulting surface is one of revolution about the $z$-axis


It is possible to define a local basis, not necessarily orthonormal with respect to the standard inner product of $\mathbb{R}^3$, as
\be
\ev_1= \frac{\pt_{\vartheta} \pv}{|\pt_{\vartheta} \pv|}, \qquad \ev_2=\frac{\pt_{\varphi} \pv}{|\pt_{\varphi} \pv|}, \qquad \bnu=\frac{\ev_1 \times \ev_2}{|\ev_1 \times \ev_2|}
\ee
The explicit expressions of these vector in terms of $r$ are given in ESM~\ref{sec:computations}. We then describe the tangent director field $\nv$ on $\esse$ by prescribing the angle $\alpha=\alpha(\vartheta,\varphi)$ between $\n$ and the local orthogonal basis on the tangent plane $(\ev_1,\ev_1^\perp)$, where $\e_1$ is defined in \eqref{eq:evevperpnormal} and $\ev_1^\perp=\normal\times\ev_1$,
\be
\label{eq:n_def}
\nv=\sin\alpha\, \ev_1+\cos\alpha\, \ev_1^\perp.
\ee

Let $\rho_0$ denote the mass density of a ground-state sphere with $r \equiv r_0$. The local mass density $\rho(\vartheta,\varphi)$ of a smooth deformation of the sphere, described by $\pv$ in \eqref{eq:pointp}, is determined by the principle of mass conservation. Following the argument in ESM~\ref{sec:mass_conservation}, it can be expressed in terms of $r$ as
\be
\label{eq:rhoformula}
\rho(\vartheta,\varphi)=\rho_0\frac{r_0^2\sin\vartheta}{r\sqrtthree}.
\ee

Furthermore, the conservation of the volume enclosed by the bubble, given by \eqref{eq:constraint}, is enforced through the following global constraint:
\be
\frac{1}{3}\int_{0}^{2\pi} \int_{0}^{\pi} r^3(\vartheta,\varphi) \sin \vartheta\; \dd  \vartheta \;\dd \varphi = V_0.
\label{eq:vol}
\ee

In ESM~\ref{sec:computations}, we provide the derivations of the operators on the surfaces $\esse$ under consideration that appear in both the energy and the equilibrium equations, expressed in terms of $r$.

We now investigate the effects of decreasing the temperature of a nematic confined within a deformable spherical shell. At sufficiently high temperatures, no preferred in-plane direction exists, and the system remains in the isotropic state, characterised by $q = 0$.

As the temperature is lowered, the nematic order emerges as soon as the temperature reaches a critical value $T_{\rm cr}$. In the present work, this onset of ordering is modelled by the Landau-de Gennes parameter $a$ in \eqref{eq:w_LdG} decreasing below a critical threshold $a_{\rm cr}$.
The onset of nematic order thus arises as a bifurcation from the trivial solution, given by $r = r_0$ and $q = 0$. This transition introduces both anisotropy, owing to the presence of a preferred local direction, and mass non-uniformity, necessitated by the topological constraints of a closed surface. Consequently, the spherical symmetry of the shell is broken, giving rise to a diverse array of possible equilibrium morphologies.

To study this bifurcation problem, we look for solutions $q$ and $r_1$ of the equilibrium equations \eqref{eq:set_equations} close to the spherical isotropic configuration, $r\equiv r_0$ and $q=0$, in the form
\begin{subequations}
\begin{align}
\label{eq:q_r_perturbed}
q(\vartheta,\varphi) &= \varepsilon q_1(\vartheta,\varphi), \\
r(\vartheta,\varphi) &= r_0 + \varepsilon r_1(\vartheta,\varphi),\label{eq:quasispherical_radius}
\end{align}
\end{subequations}
where $\varepsilon \ll 1$ is a dimensionless parameter which provides a measure of small departures of $a$ from its value $a_{\rm cr}$ at the nematic--isotropic transition temperature $T_{\rm cr}$ on the sphere. 

We will be more precise about how $\varepsilon$ is related to $a$ and $a_{\rm cr}$ in Section~\ref{sec:weak_nonlinear}, where we focus on the nature of the bifurcation with respect to $a$, identifying in which cases it is of first or second order.

In ESM~\ref{sec:expansionvarepsilon}, we present the first-order expansion in $\varepsilon$ of the quantities involved in the energy and in the equilibrium equations.

The specialised forms of equations \eqref{eq:n2} and \eqref{eq:q2} for a quasi-spherical surface are presented below. In the subsequent expressions, the differential operators $\Deltastar_s$ and $\gradstar$ are understood to denote the Laplace–Beltrami operator and the surface gradient, respectively, on a spherical surface of radius $r_0$. 

At leading order, $O(\varepsilon^2)$, equation \eqref{eq:n2} reduces to:
\begin{align}
4 k \varepsilon^2 \rho_0 \left[ q_1^2 \Deltastar_s \alpha + 2 q_1 \left( \gradstar \alpha \cdot \gradstar  q_1 - \frac{\cot \vartheta}{r_0^2 \sin \vartheta} \frac{\pt q_1}{\pt \varphi} \right) \right]\nonumber\\ - \eeta
 \frac{4 k \varepsilon^2 \rho_0 q_1}{r_0^3} \left( \sin(2\alpha) \left[ r_0^2 \Deltastar_s r_1 - 2\frac{\pt^2 r_1}{\pt \vartheta^2} \right] + 2\cos(2\alpha) \csc \vartheta \left[\cot \vartheta\frac{\pt r_1}{\pt \varphi} - \frac{\pt^2 r_1}{\pt \varphi \pt \vartheta}\right] \right) = 0,
 \label{eq:nlin}
\end{align}
whereas at $O(\varepsilon)$, equation \eqref{eq:q2} yields
\begin{align}
 \varepsilon\rho_0 \left[k \Deltastar_s q_1 - q_1 \left( a - \frac{2k}{r_0^2}\right) - 4k q_1 \left(|\gradstar \alpha|^2 +   \frac{\csc^2 \vartheta}{r_0^2}  - \frac{2}{r_0^2}\cot \vartheta \csc \vartheta \frac{\pt \alpha}{\pt \varphi} \right) \right] \nonumber\\
 + \frac{2k \varepsilon \rho_0}{r_0^3} \left[ \cos(2\alpha) \left( r_0^2 \Deltastar_s r_1 - 2 \frac{\pt^2 r_1}{\pt \vartheta^2} \right) - 2\sin(2\alpha) \csc \vartheta \left( \cot \vartheta \frac{\pt r_1}{\pt \varphi} - \frac{\pt^2 r_1}{\pt \varphi \pt \vartheta}  \right) \right]=0.
 \label{eq:qlin}
\end{align}

These two lengthy equations can be condensed into a single compact expression by introducing the complex variable $\QQ=q_1 {\rm e}^{2 i \alpha}$. A direct calculation thus yields the equation
\be
\frac{k}{r_0^2} \eth_1(\bar \eth_2 \QQ) - a \QQ - 2\frac{k}{r_0^3} \eth_1(\eth_0 r_1) = 0
\label{eqQ}
\ee
where
\be
\eth_\sigma = \pt_{\vartheta} - i \csc \vartheta \pt_{\varphi} - \sigma \cot \vartheta, 
\qquad
\bar \eth_\sigma = \pt_{\vartheta} + i \csc \vartheta \pt_{\varphi} + \sigma \cot \vartheta,
\label{eq:eth}
\ee
are the spin-raising and spin-lowering operator, respectively \cite{Goldberg:1967}.

By combining all the resulting terms at order $O(1)$, we obtain the Young--Laplace law~\eqref{eq:YL}.

At order $O(\varepsilon)$, the introduction of the complex variable $\QQ=q_1 \mathrm{e}^{2 \mathrm{i} \alpha}$ once again yields a compact representation:
\be
- \left(\Delta^{0}_s + \frac{2}{r_0^2}\right) \Big(  \gamma \, r_1 + \frac{k}{r_0} \, \rho_1 \Big) + \frac{k \rho_0}{r_0^2} \left(\Delta^{0}_s + \frac{3}{r_0^2}\right)\left(\Delta^{0}_s + \frac{2}{r_0^2}\right) r_1 + 2 \frac{k \rho_0}{r_0^3} \mathrm{Re}[\bar\eth_1(\bar\eth_2 \QQ)] - \mu_1 = 0.
\label{eq:sh1}
\ee

Since mass is conserved, equation~\eqref{eq:rhoformula} must hold. At first order, this yields
\be
\rho_1 = -2 \rho_0 \frac{r_1}{r_0},
\ee
which, when substituted into~\eqref{eq:sh1}, leads to
\begin{equation}
\label{eq:r1EL}
\left[-\gamma+k\rho_0\left(\Deltastar_s+\frac{5}{r_0^2}\right)\right]\left(\Deltastar_s+\frac{2}{r_0^2}\right)r_1 + 2 \frac{k \rho_0}{r_0^3} \mathrm{Re}[\bar\eth_1(\bar\eth_2 \QQ)] -  \mu_1 = 0.
\end{equation}

Furthermore, volume conservation imposes the constraint
\be
\int_{0}^{2\pi} \int_{0}^{\pi} r_1(\vartheta,\varphi) \sin \vartheta \; \dd \vartheta \; \dd \varphi = 0.
\label{eq:volume1}
\ee

Equations~\eqref{eqQ}, \eqref{eq:r1EL} and \eqref{eq:volume1} constitute a system of integro-differential equations for the unknowns $\QQ$, $r_1$ and $\mu_1$, which is trivially satisfied when all these fields vanish. However, as our interest lies in non-trivial solutions, solving this system must be framed as an eigenvalue problem, where the eigenvalues correspond to specific values of the parameter $a$.

\section{Bifurcation analysis}

\subsection{Linear analysis}
We aim to determine the eigenvalues $a$ and the associated eigenfunctions for the fields $\QQ$ (with spin-weight $s=2$) and $r_1$ (with spin-weight $s=0$), alongside the constant $\mu_1$, governed by the system of equations~\eqref{eqQ}, ~\eqref{eq:r1EL} and \eqref{eq:volume1}.

We expand the unknown functions in terms of the spin-weighted spherical harmonics ${}_\sigma Y_{lm}(\vartheta,\varphi)$, which form a complete basis and are eigenfunctions of the relevant angular differential operators. Thus,
\begin{equation}
\label{eq:eigenfunctions_general}
\QQ(\vartheta,\varphi) = \sum_{l,m} Q_{lm} \, {_2Y_{lm}}(\vartheta,\varphi), \qquad 
r_1(\vartheta,\varphi) = \sum_{l,m} R_{lm} \, Y_{lm}(\vartheta,\varphi),
\end{equation}
where $Q_{lm}$ and $R_{lm}$ denote the expansion coefficients.

The actions of the standard spin-raising and spin-lowering operators on the basis functions are well known:
\begin{align}
\eth_\sigma({_sY_{lm}}) &= \sqrt{(l-\sigma)(l+\sigma+1)} \, {_{\sigma+1}Y_{lm}}, \\
\bar{\eth}_\sigma({_\sigma Y_{lm}}) &= -\sqrt{(l+\sigma)(l-\sigma+1)} \, {_{\sigma-1}Y_{lm}}, \\
\Deltastar_s(Y_{lm}) &= -\frac{l(l+1)}{r_0^2} \, Y_{lm}.
\end{align}

Substituting the eigenfunction expansion \eqref{eq:eigenfunctions_general} into Eqs. \eqref{eqQ} and \eqref{eq:r1EL}, we observe that $\mu_1$ is non-vanishing only for the $l=0$ mode. In this case, Eq. \eqref{eqQ} is trivially satisfied, while Eq. \eqref{eq:r1EL} reduces to
\be
\frac{2}{r_0^2} \left[ -\gamma + \frac{5 k \rho_0}{r_0^2} \right] R_{00} - \mu_1 = 0.
\ee
The constraint \eqref{eq:volume1} implies $R_{00} = 0$, which consequently leads to $\mu_1 = 0$.

For each mode with $l\geq 1$ and $-l\le m \le l$, equations \eqref{eqQ} and \eqref{eq:r1EL} yield
\begin{subequations}
\begin{align}
\left(-\frac{k}{r_0^2}(l-1)(l+2) - a\right) Q_{lm} 
- \frac{2k}{r_0^3} \sqrt{l(l+1)(l-1)(l+2)} \, R_{lm} & = 0, \\
\left(-\gamma+\frac{k\rho_0\left(5-l(l+1)\right)}{r_0^2}\right)R_{lm} 
+ \frac{k\rho_0}{r_0}\frac{\sqrt{l(l+1)(l-1)(l+2)}}{\left(2-l(l+1)\right)} \, (Q_{lm} + (-1)^m Q_{l\; -m}^{0})&= 0.
\label{sys1}
\end{align}
\end{subequations}
which is non-trivial only for $l \ge 2$. Moreover, each of these modes satisfies the constraint~\eqref{eq:volume1}.

It should be noted that only when
\be
Q_{lm} = (-1)^m Q_{l,-m}^*
\label{eq:reality}
\ee
holds can Eq.~\eqref{sys1} be recast as a homogeneous linear system for the variables $(Q_{lm}, R_{lm})$:
\be
\label{eq:matrix}
\begin{pmatrix}
\dfrac{k\rho_0(l^2 + l - 2)}{r_0^2} + a\rho_0 &
\dfrac{2k\rho_0\sqrt{l(l+1)(l^2 + l - 2)}}{r_0^3} \\[8pt]
\dfrac{2k\rho_0\sqrt{l(l+1)(l^2 + l - 2)}}{r_0^3} &
-\dfrac{(l^2 + l - 2)}{r_0^2}\left(-\gamma+\dfrac{k\rho_0\left(5-l(l+1)\right)}{r_0^2}\right)
\end{pmatrix}
\begin{pmatrix}
Q_{lm} \\[3pt]
R_{lm}
\end{pmatrix}
=
\begin{pmatrix}
0 \\[3pt]
0
\end{pmatrix},
\ee
and, hence, the existence of non-trivial solutions requires the determinant of the coefficient matrix to vanish, which yields the condition
\be
a_l = - \frac{k(l^2+l-2)}{r_0^2} - \frac{4 k^2 \rho_0\, l(l+1)}{r_0^4 \left(-\gamma + \dfrac{k \rho_0 (5-l(l+1))}{r_0^2}\right)}.
\label{eq:al}
\ee
Furthermore, solving system \eqref{eq:matrix} yields the amplitude ratio:
\begin{equation}
\Lambda_l:=\frac{Q_{lm}}{R_{lm}}
= -\frac{2k\sqrt{l(l+1)(l-1)(l+2)}/r_0^3}{a_l + k(l-1)(l+2)/r_0^2}.
\end{equation}

The critical bifurcation threshold corresponds to the maximum value of $a$ within the spectrum \eqref{eq:al}, which occurs for the $l=2$ mode:
\be
\label{eq:a2def}
a_{\rm cr} := a_{2}=- \frac{2 k}{r_0^2} + \frac{24 k^2 \rho_0}{r_0^4 \left(\gamma + \frac{k \rho_0}{2r_0^2}\right)},
\ee
with the associated amplitude ratio:
\be
\Lambda_{\rm cr}\;=-\frac{1}{\sqrt{6}}\left(\frac{2\gamma r_0}{k\rho_0}+\frac{1}{r_0}\right).
\ee
The explicit expressions for the spin-weight 0 (standard) and spin-weight 2 spherical harmonics with $l=2$ are listed in Table \ref{tab:modes} in ESM. The reality condition \eqref{eq:reality} necessitates the combination of modes with opposite $m$, restricting the admissible $l=2$ eigenfunctions to the three presented in Table \ref{tab:modes_constrained}.
\begin{table}[h]
\centering
\renewcommand{\arraystretch}{1.8}
\begin{tabular}{@{}cccc@{}}
\hline
$m$ & \text{ $r_1$ eigenfunctions} & \text{$\QQ$ eigenfunctions} \\ \hline
$-2,2$ & $ \sqrt{\frac{15}{8\pi}} \sin^2 \vartheta  \, \left(\mathrm{Re}[Q_{22}]\cos 2\varphi-\mathrm{Im}[Q_{22}]\sin2\varphi\right) $ & $ \frac{1}{2}\sqrt{\frac{5}{\pi}} \left(Q_{22} \text{e}^{ 2 i \varphi } \cos^4\frac{\vartheta}{2}+Q_{22}^* \text{e}^{ -2 i u } \sin^4\frac{\vartheta}{2}\right)$ \\
$-1,1$ & $ \sqrt{\frac{15}{8\pi}} \sin 2\vartheta\left(-\mathrm{Re}[Q_{21}]\cos \varphi+\mathrm{Im}[Q_{21}]\sin \varphi\right)$ & $ \sqrt{\frac{5}{\pi}} \sin \frac{\vartheta}{2}\cos\frac{\vartheta}{2} \left(Q_{21} \text{e}^{i \varphi}\cos^2\frac{\vartheta}{2}-Q_{21}^* \text{e}^{-i \varphi}\sin^2\frac{\vartheta}{2}\right) $ \\
$0$ & $ \frac{1}{4}\sqrt{\frac{5}{\pi}} \mathrm{Re}[Q_{20}](3\cos^2 \vartheta  - 1) $ & $ \frac{1}{4}\sqrt{\frac{15}{2\pi}} \mathrm{Re}[Q_{20}]\sin^2 \vartheta  $ \\ \hline
\end{tabular}
\caption{Eigenfunctions associated with the modes $l=2$ and $m=-2, \, -1, \, 0, \, 1, \, 2$, under the constraints in \eqref{eq:reality}.}
\label{tab:modes_constrained}
\end{table}

\subsection{Weakly nonlinear analysis}\label{sec:weak_nonlinear}
As the temperature decreases, the parameter $a$ diminishes accordingly. Once $a$ reaches the critical value $a_{\rm cr}$ defined in \eqref{eq:a2def}, the ground-state solution becomes unstable. The onset of nematic ordering, and the consequent loss of spherical symmetry, therefore occurs via a bifurcation from this trivial configuration.

It is fundamental to ascertain whether this transition is continuous (second-order) or discontinuous (first-order) with respect to $a$. These transitions are governed by distinct physical mechanisms~\cite{landau_lifshitz:1980, strogatz:2014}: a continuous transition is inherently reversible, retracing the same family of solutions if $a$ is reversed, whereas a first-order transition may exhibit hysteresis. In the latter case, the solution follows a different branch, preventing a return to the initial state merely by reversing the process.

To analytically describe the order of the transition, we consider small departures of $a$ from $a_{\mathrm{cr}}$, expressed as
\begin{equation}
\label{eq:a_formula}
a = a_{\mathrm{cr}}(1 + \varepsilon^\lambda) < 0,
\end{equation}
where $\lambda \in \{1,2\}$. If $\lambda=1$, the bifurcation is \emph{transcritical}; otherwise, it is either \emph{supercritical} or \emph{subcritical}.

To determine the nature of the bifurcation, we analyse the free-energy functional for $a$ slightly below $a_{\mathrm{cr}}$. In this regime, the linearised solutions for $r_1$ and $\QQ$ take the forms detailed in Table~\ref{tab:modes_constrained}. Because linear theory defines each bifurcation mode only up to a multiplicative constant, we employ a weakly non-linear analysis to both characterise the bifurcation type and determine the mode amplitudes.

Following Koiter's asymptotic approach~\cite{koiter:1981}, we substitute the linear eigenmodes for $r_1$ and $\QQ$ (Table~\ref{tab:modes_constrained}) alongside \eqref{eq:a_formula} into the free-energy functional, expanding it in powers of $\varepsilon$ up to the fourth order. The functional thus reduces to a function of the complex amplitudes $Q_{2m}$, with $m\in\{0,1,2\}$:
\begin{equation}
\label{eq:thirdfourthorder}
\mathcal{F}\left[Q_{20},\, Q_{21}, \, Q_{22}\right]
=\varepsilon^{3} W^{(3)}\left[Q_{20},\, Q_{21}, \, Q_{22}\right]
+ \varepsilon^{4} W^{(4)}\left[Q_{20},\, Q_{21}, \, Q_{22}\right]
+ O(\varepsilon^{5}),
\end{equation}
where $W^{(3)}$ and $W^{(4)}$ represent the cubic and quartic terms, respectively. The zeroth-order term is constant, whilst the first- and second-order terms vanish. Selecting the scaling exponent $\lambda \in \{1, 2\}$ is crucial and dictated by the presence or absence of the cubic term. The equilibrium values of $Q_{2m}$ are then determined as stationary points of $\mathcal{F}$. ESM~\ref{sec:modalanalysis_appendix} details the derivation of the resulting real-valued function.

\subsubsection{Axisymmetric configurations}\label{sec:lin_axisym}
Focusing on the azimuthal mode $m=0$ (Table~\ref{tab:modes_constrained}), the resulting equilibrium configuration for the radius $r$, defined in \eqref{eq:quasispherical_radius} and given by
\be
\label{eq:rm0_def}
r^{m=0}(\vartheta) = r_0 + \varepsilon r_{1,20}(\vartheta),
\ee
describes a surface of revolution, as the perturbation $r_{1,20}$ is independent of the azimuthal angle $\varphi$. The associated nematic texture exhibits two antipodal melting points at the poles, where the order parameter $q_1$ vanishes, each carrying a topological charge of $+1$.

Defining the real amplitude $C_0 := \mathrm{Re}[Q_{20}]$ and introducing the dimensionless parameters
\begin{equation}
\label{eq:gamma0_def}
\bar \gamma := \frac{2\gamma r_0^2}{k\rho_0},
\qquad
\bar{c} := \frac{cr_0^2}{k},
\end{equation}
we take $\lambda=1$ and identify the following expansion coefficients:
\begin{subequations}
\begin{gather}
\label{eq:W3W4C0}
W^{(3)}[C_0] = -\frac{k \rho_0 C_0^2 \left[ -14 (-5 + \bar \gamma) (1 + \bar \gamma)^2 + (-41 - 10 \bar \gamma + 11 \bar \gamma^2) \sqrt{\frac{30}{\pi}} \, C_0 \right]}{14 (1 + \bar \gamma)^3}, \\
W^{(4)}[C_0] = \frac{5 C_0^4 \left( \bar{c} (1 + \bar \gamma)^4 + 54 (119 + 28 \bar \gamma + \bar \gamma^2) \right) k \rho_0}
{56 (1 + \bar \gamma)^4 \pi}.
\end{gather}
\end{subequations}
Since $W^{(3)} \neq 0$, the bifurcation is transcritical; hence, both terms contribute to the local minimisation conditions. Minimising $\mathcal{F}$ with respect to $C_0$ yields the trivial ground-state solution:
\begin{equation}
\label{eq:c0_isotropy}
C_0=0,
\end{equation}
which holds for all $\bar \gamma>0$, alongside the bifurcated anisotropic solutions:
\begin{subequations}
\begin{gather}
\label{eq:c0_anisotropy}
C_{0,\pm}^{\bar \gamma}=\frac{\sqrt{\pi}(1 + \bar \gamma)}{
\varepsilon\sqrt{10}\left( \bar{c} (1 + \bar \gamma)^4 + 54 (119 + 28 \bar \gamma + \bar \gamma^2) \right)}
\left[123 \sqrt{3}+ 30 \sqrt{3}\, \bar \gamma- 33 \sqrt{3}\, \bar \gamma^2\right.\nonumber\\
\left.\pm \sqrt{27 (41 + 10 \bar \gamma - 11 \bar \gamma^2)^2
+ 56 \varepsilon(-5 - 4 \bar \gamma + \bar \gamma^2)
\left( \bar{c} (1 + \bar \gamma)^4 + 54 (119 + 28 \bar \gamma + \bar \gamma^2) \right)}\right].
\end{gather}
\end{subequations}
These solutions exist when the radicand is non-negative, are distinct when it is strictly positive, and coincide when it vanishes.

In the asymptotic limit of large $\bar \gamma$, \eqref{eq:c0_anisotropy} simplifies to:
\begin{equation}
\label{eq:c0_asymptotic_limit}
C_{0,\pm}^{\bar \gamma} = \pm \sqrt{\frac{28\pi}{5\bar c\varepsilon}} \left[ 1 - \left( \pm \frac{33\sqrt{42}}{28\sqrt{\bar c\varepsilon}} + 3 \right) \frac{1}{\bar \gamma} + \mathcal{O}\left(\frac{1}{\bar \gamma^2}\right) \right].
\end{equation}

\begin{figure}[t]
    \centering
    \begin{subfigure}{0.32\textwidth}
        \centering
        \vspace{0.8cm}
        \includegraphics[width=\textwidth]{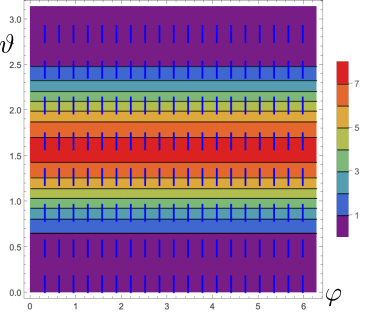}
        \caption{Contour plot of $q_{1,20}=|Q_{20}|$ for the prolate shape ($C_{0,\pm}^{\bar \gamma}<0$).} 
        \label{fig:countorplotc0minus}
    \end{subfigure}  
    \begin{subfigure}{0.14\textwidth}
        \centering
        \includegraphics[width=\textwidth]{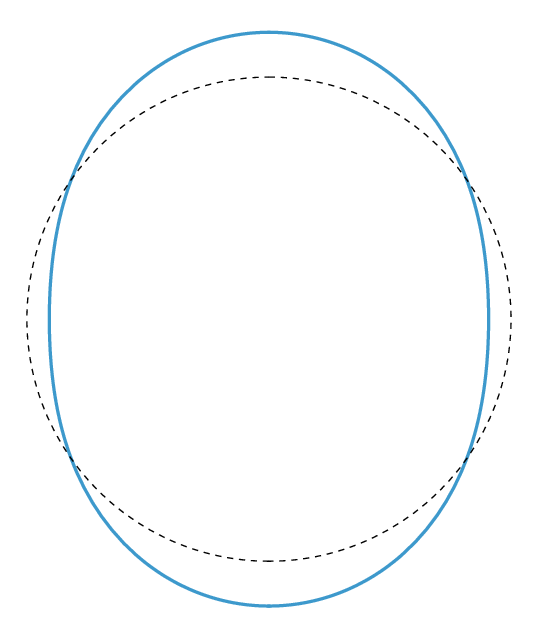}
        \includegraphics[width=\textwidth]{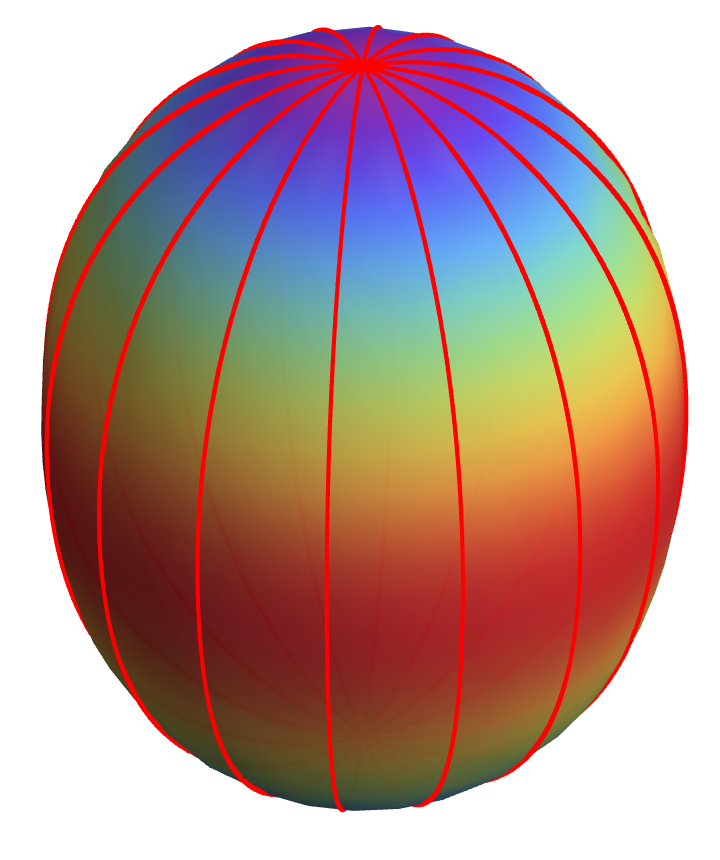}
        \vspace{0.3cm}
    \end{subfigure}
    \hspace{0.4cm}
    \begin{subfigure}{0.32\textwidth}
        \centering
        \vspace{0.8cm}
        \includegraphics[width=\textwidth]{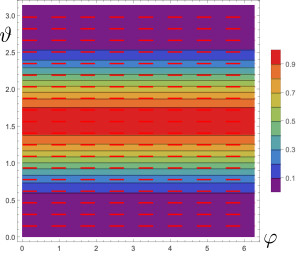}
        \caption{Contour plot of $q_{1,20}=|Q_{20}|$ for the oblate shape ($C_{0,+}^{\bar \gamma}>0$).} 
        \label{fig:countorplotc0plus}
    \end{subfigure}  
    \begin{subfigure}{0.14\textwidth}
        \centering
        \includegraphics[width=\textwidth]{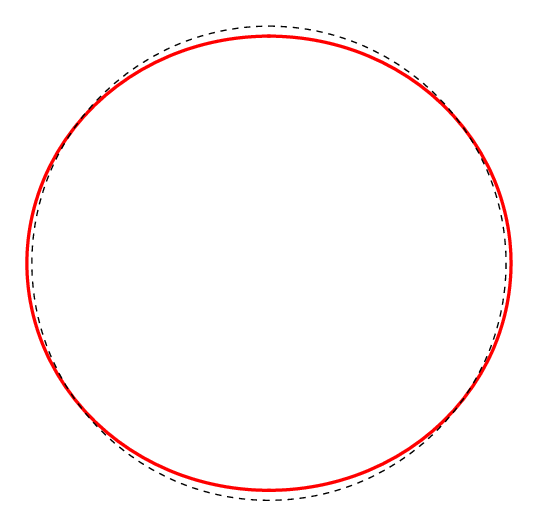}
        \includegraphics[width=\textwidth]{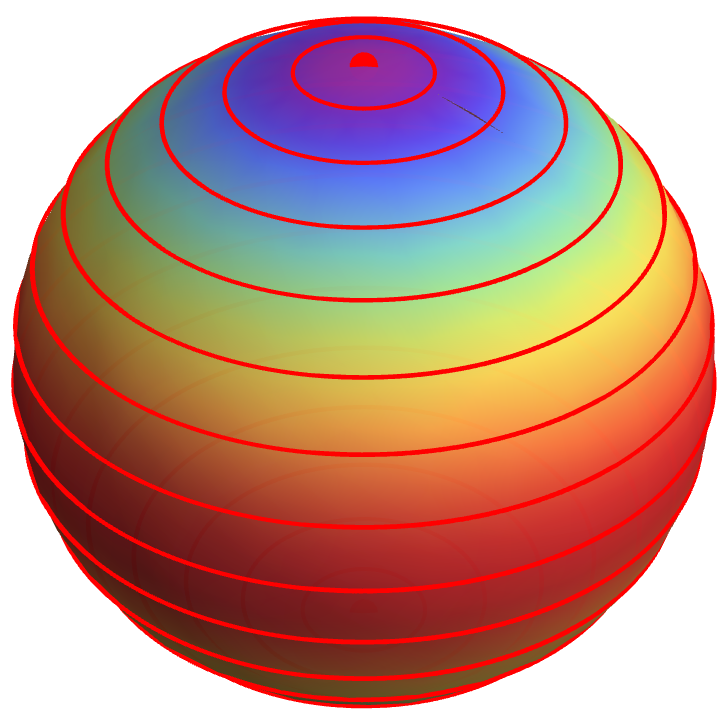}
        \vspace{0.5cm}  
    \end{subfigure}
    \caption{Nematic textures corresponding to the bifurcated equilibrium configurations for prolate (Left) and oblate (Right) shapes. Two $+1$ point defects are situated at the poles ($\vartheta = 0$ and $\vartheta = \pi$). The lateral insets illustrate the meridional cross-sections and three-dimensional renderings of the deformed shells alongside their associated nematic textures. Within the cross-sectional views, black dashed lines denote the reference sphere.}
    \label{fig:countorplots_c0}
\end{figure}

\begin{figure}[t]
    \centering
    \begin{subfigure}[c]{0.49\linewidth}
        \centering
        \includegraphics[width=0.7\linewidth]{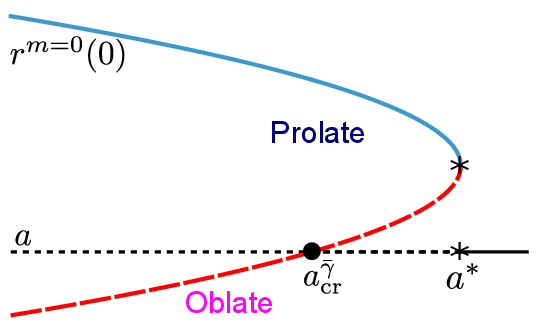}
        \caption{Equilibrium values of $r^{m=0}$ in \eqref{eq:rm0_def} at the pole $\vartheta = 0$ vs $a$. The horizontal line shows the trivial isotropic solution $r_0$. A fold at $\astar > a_{\rm cr}^{\bar \gamma}$ gives two non-trivial equilibrium solutions.}
        \label{fig:r_epsilon0}
    \end{subfigure}
    \hspace{0.2cm}
    \begin{subfigure}[c]{0.45\linewidth}
        \centering
        \includegraphics[width=0.8\linewidth]{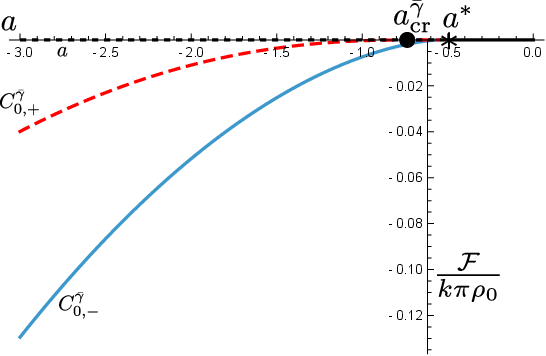}
        \caption{Graph of the dimensionless free-energies $\mathcal{F}[C_{0,+}^{\bar \gamma}]/(k\pi\rho_0)$ (red), $\mathcal{F}[C_{0,-}^{\bar \gamma}]/(k\pi\rho_0)$ (blue), and $\mathcal{F}[0]$ (black) as functions of $a$. The blue branch has lower free energy.}
        \label{fig:F_epsilon0}
    \end{subfigure}
    \caption{Overview of the first-order bifurcation scenario. The blue and red lines refer to the equilibrium branches corresponding to $C_{0,-}^{\bar \gamma}$ and $C_{0,+}^{\bar \gamma}$, respectively, while the black line refers to the isotropic trivial solution $C_0 \equiv 0$. Dashed lines indicate metastable solutions, and solid lines indicate stable ones. Here, $\bar \gamma=8$, $k=1$, $\rho_0=0.1$, $r_0=1$, and $\bar{c}=2$. Consequently, $a_{\mathrm{cr}}^{\bar \gamma=8}=-0.67$, while $\astar=-0.5$.}
    \label{fig:rF_epsilon0}
\end{figure}

To visualise the symmetry-breaking behaviour of this transcritical bifurcation, Figure~\ref{fig:r_epsilon0} illustrates the dependence of the equilibrium solution $r^{m=0}$ (evaluated at the pole $\vartheta=0$) upon the parameter $a$ near $a_{\mathrm{cr}}$. We consider the radial profiles $r_{1,20}$ associated with the amplitudes $C_0$ in \eqref{eq:c0_isotropy} and \eqref{eq:c0_anisotropy}. We express $\varepsilon$ in terms of $a$ using \eqref{eq:a_formula} with $\lambda = 1$:
\begin{equation}
\label{eq:varepsilon_lambda1}
\varepsilon = \frac{a - a_{\mathrm{cr}}^{\bar \gamma}}{a_{\mathrm{cr}}^{\bar \gamma}},
\end{equation}
where $a_{\mathrm{cr}}^{\bar \gamma}$ is recast as a function of $\bar \gamma$:
\begin{equation}
\label{eq:aval2scaled}
a_{\mathrm{cr}}^{\bar \gamma} = -\frac{2k}{r_0^2} \left( 1 - \frac{6}{1+\bar \gamma} \right).
\end{equation}

The sign of $C_{0,\pm}^{\bar \gamma}$ governs both the morphology of the bifurcated shape and the nematic orientation. Note that $C_{0,+}^{\bar{\gamma}}$ is negative for $a < a_{\mathrm{cr}}^{\bar{\gamma}}$, vanishes at $a = a_{\mathrm{cr}}^{\bar{\gamma}}$, and becomes positive for $a > a_{\mathrm{cr}}^{\bar{\gamma}}$, whereas $C_{0,-}^{\bar{\gamma}}$ remains strictly positive for all admissible values of $a$. Specifically, a negative sign yields prolate shapes (elongated along the polar axis), because the shape perturbation amplitude involving $C_{0,\pm}^{\bar \gamma}/\Lambda_{\mathrm{cr}}$ ($\Lambda_{\mathrm{cr}}<0$) is positive. The amplitude of $\QQ$ is real and negative, implying the nematic director aligns along the meridians. Conversely, a positive $C_{0,\pm}^{\bar \gamma}$ yields oblate shapes, with the nematic director aligned along the parallels.

Figure~\ref{fig:countorplots_c0} illustrates the prolate and oblate configurations alongside their nematic textures, presented as 3D axisymmetric surfaces and as contour plots in the $(\varphi, \vartheta)$ plane. The surface colour map denotes the modulus $q_{1,20}=|Q_{20}|$, with the darkest regions indicating the topological defects. The blue and red curves depict the integral lines of the nematic director for the prolate and oblate geometries, respectively. In both configurations, two point defects invariably locate at the poles ($\vartheta = 0$ and $\vartheta = \pi$).

Figure~\ref{fig:r_epsilon0} illustrates the bifurcation diagram for the axisymmetric mode. The horizontal line $r \equiv r_0$ denotes the trivial spherical solution, valid for all values of $a$. During an idealised cooling process ($a$ decreases), a fold (or limit point) emerges at $\astar > a_{\mathrm{cr}}$, where $\astar$ is the value of $a$ for which the discriminant in \eqref{eq:c0_anisotropy} vanishes:
\begin{equation}
\label{eq:astar_def}
\astar:=\frac{-56 k\bar c (-5 + \bar \gamma) (1 + \bar \gamma)^5 + 
     27 (68321 + 69812 \bar \gamma - \bar \gamma^2 - 2908 \bar \gamma^3 + 
        9 \bar \gamma^4)}{28 (1 + \bar \gamma)^2 (\bar c (1 + \bar \gamma)^4 + 
     54 (119 + 28 \bar \gamma + \bar \gamma^2))r_0^2}.
\end{equation}
Consequently, the ground-state solution ceases to be the unique equilibrium well before $a$ reaches $a_{\mathrm{cr}}$. These emergent states correspond to $C_{0,-}^{\bar \gamma}$ and $C_{0,+}^{\bar \gamma}$, yielding the blue and red branches, respectively. The system exhibits symmetry breaking: the branches of $r^{m=0}$ lack reflectional symmetry about the horizontal axis ($C_{0,+}^{\bar \gamma} \neq - C_{0,-}^{\bar \gamma}$).

To complement the bifurcation diagram, we evaluate the energy along each branch. Substituting $C_{0,\pm}^{\bar \gamma}$ and $\varepsilon$ from \eqref{eq:varepsilon_lambda1} into $\mathcal{F}$ reveals that:
\begin{equation}
\label{eq:energy_inequality}
\mathcal{F}[C_{0,-}^{\bar \gamma}](a) < \mathcal{F}[C_{0,+}^{\bar \gamma}](a) < 0
\end{equation}
holds whenever the non-trivial solutions exist ($a < \astar$). The corresponding values are plotted in Figure~\ref{fig:F_epsilon0}. 

Adding this information to Figure~\ref{fig:r_epsilon0}, solid lines denote global minima and dashed lines indicate local (metastable) minima. For $a > \astar$, $\mathcal{F}$ attains a single, globally stable minimum at $r^{m=0}\equiv r_0$. The scenario alters at the fold point $a =\astar$, where two non-trivial solutions emerge. The blue branch ($C_{0,-}^{\bar \gamma}$) immediately becomes the globally stable equilibrium, supplanting the isotropic branch. The red branch ($C_{0,+}^{\bar \gamma}$) remains metastable for all $a < \astar$, despite possessing a lower energy than the isotropic state.

The trivial solution therefore becomes metastable at $a = \astar$, rather than at $a_{\mathrm{cr}}^{\bar \gamma}$ where the red branch intersects the isotropic one. Because the blue branch exhibits a lower free energy for all $a < \astar$, the system undergoes a \emph{discontinuous} first-order phase transition to the prolate configuration. During cycles of cooling and heating, transcritical transitions of this nature can consequently give rise to hysteresis.

To facilitate a rigorous comparison with the rigid sphere model~\cite{Napoli:2021}, we consider the asymptotic limit $\bar \gamma \to +\infty$. Here, $\mathcal{F}[C_{0,\pm}^{\bar \gamma}]$ reduces to:
\begin{equation}
\label{eq:free_energyinfinityC0}
\mathcal{F}[C_{0,\pm}^{\bar \gamma}]
\overset{\bar \gamma\to+\infty}{\longrightarrow}
-\frac{14 k \pi \rho_0}{5\bar{c}}
\left(\frac{a - a_{\mathrm{cr}}^\infty}{a_{\mathrm{cr}}^\infty}\right)^2, \qquad a_{\mathrm{cr}}^\infty=-\frac{2k}{r_0^2}.
\end{equation}
This perfectly recovers the result obtained in~\cite{Napoli:2021} for a rigid shell. Evaluating the limit $\bar \gamma \to +\infty$ in \eqref{eq:c0_anisotropy} retrieves the previously derived equilibrium amplitudes:
\begin{equation}
\label{eq:c0_infinity}
C_{0,\pm}^{\infty} = \pm \sqrt{ \frac{28 \pi a_{\mathrm{cr}}^\infty} {5\bar{c}(a - a_{\mathrm{cr}}^\infty)} }, 
\qquad 
{\astar}^{\infty} = a_{\mathrm{cr}}^\infty,
\end{equation}
demonstrating that, in this infinite-stiffness limit, the bifurcation transitions from transcritical to supercritical.

\subsubsection{Squared configurations}\label{sec:squared}
These non-axisymmetric configurations emerge for modes $m = 1$ and $m = 2$ (Table~\ref{tab:modes_constrained}). The equilibrium radii, defined by
\begin{equation}
\label{eq:rm12_def}
r^{m=1}(\vartheta,\varphi) = r_0 + \varepsilon r_{1,21}(\vartheta,\varphi), \qquad r^{m=2}(\vartheta,\varphi) = r_0 + \varepsilon r_{1,22}(\vartheta,\varphi),
\end{equation}
clearly do not constitute surfaces of revolution.

\begin{figure}[t]
    \centering
    \begin{subfigure}{0.32\textwidth}
        \centering
        \vspace{0.8cm}
        \includegraphics[width=\textwidth]{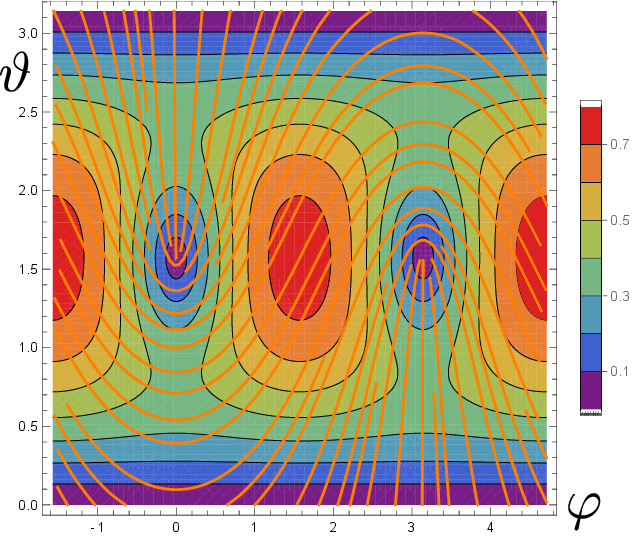}
        \caption{Contour plot of $q_{1,21}=|Q_{21}|$ for $C_{1,-}^{\bar \gamma}$.} 
        \label{fig:contourplotc1minus}
    \end{subfigure}  
    \begin{subfigure}{0.14\textwidth}
        \centering
        \includegraphics[width=\textwidth]{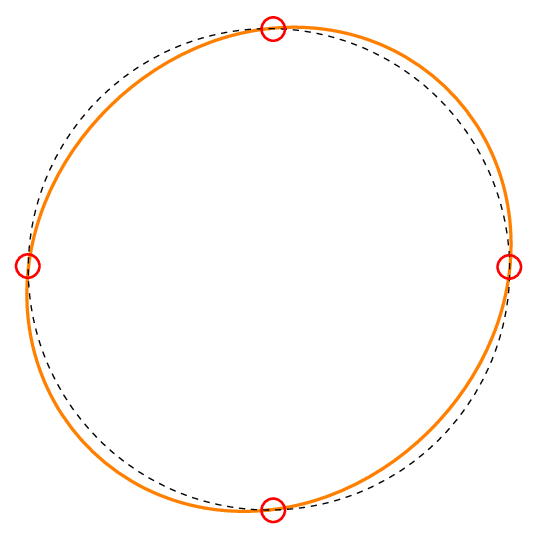}
        \includegraphics[width=\textwidth]{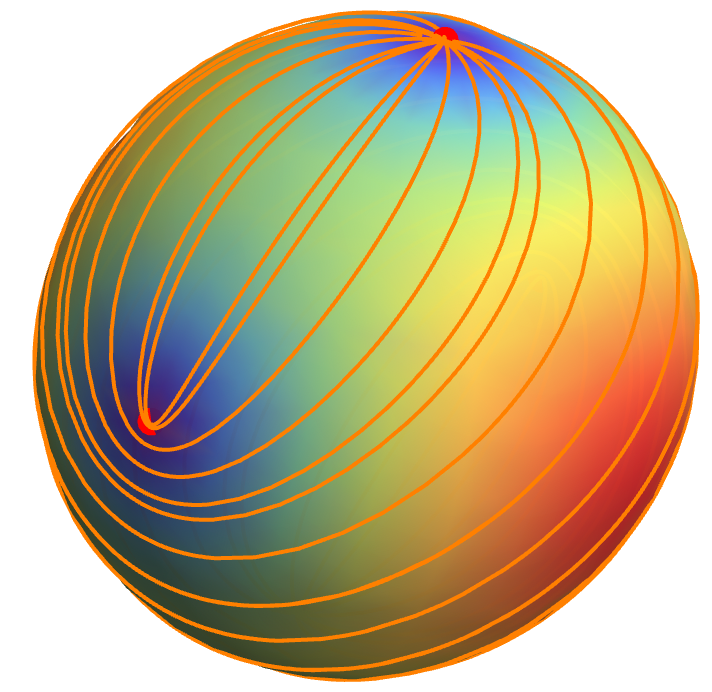}
        \vspace{0.3cm}
    \end{subfigure}
    \hspace{0.4cm}
    \begin{subfigure}{0.32\textwidth}
        \centering
        \vspace{0.8cm}
        \includegraphics[width=\textwidth]{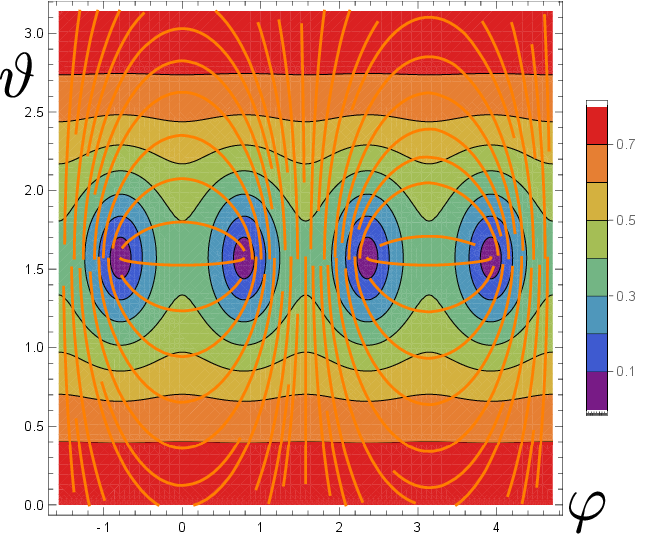}
        \caption{Contour plot of $q_{1,22}=|Q_{22}|$ for $C_{2,-}^{\bar \gamma}$.} 
        \label{fig:countorplotc0plus}
    \end{subfigure}  
    \begin{subfigure}{0.14\textwidth}
        \centering
        \includegraphics[width=\textwidth]{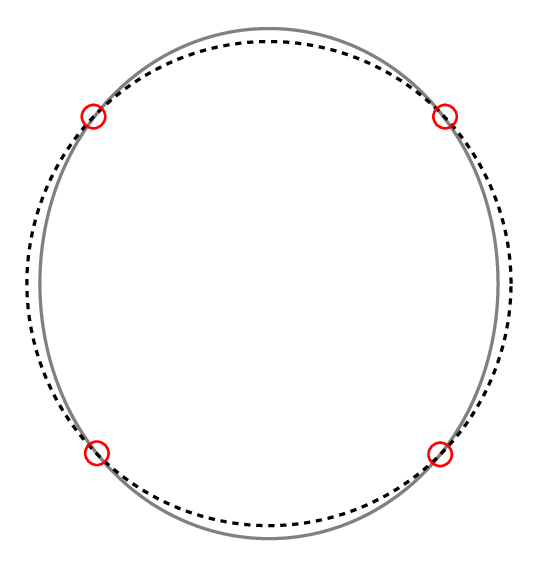}
        \includegraphics[width=\textwidth]{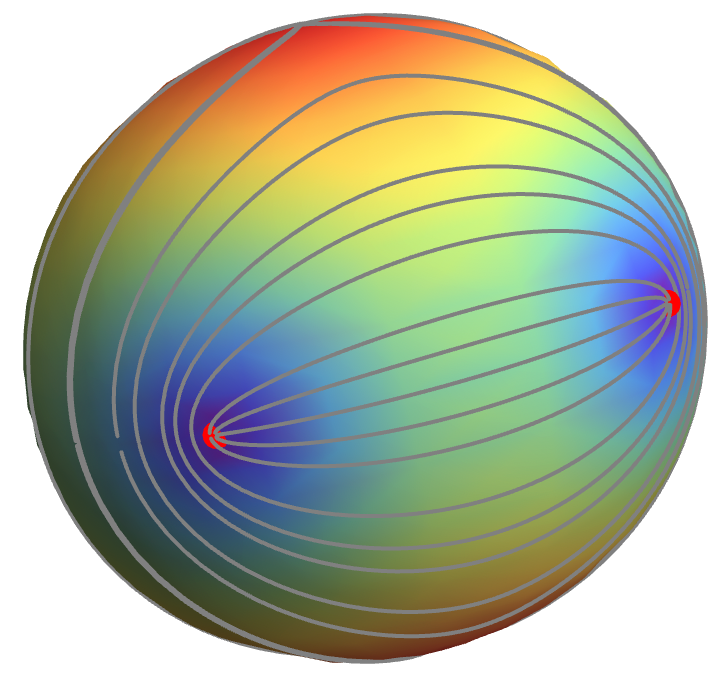}
        \vspace{0.10cm}  
     \end{subfigure}
    \caption{3D representations of the deformed shells and their corresponding nematic textures, illustrating the bifurcated equilibrium modes for $m=1$ and $m=2$. 
    In the left-hand panel ($m=1$), two defects reside on the equator whilst the remaining two are situated at the poles; conversely, in the right-hand panel ($m=2$), all four defects lie along the equator. Furthermore, the configuration of the integral lines reveals that these defects carry a topological charge of $+1/2$. In the cross-sectional views taken along planes containing the defects, black dashed lines denote the reference sphere.}
    \label{fig:contourplots_c12}
\end{figure}

As illustrated in Figure~\ref{fig:contourplots_c12}, there exist four points at the vertices of a square where $\QQ$ vanishes; we thus designate them \emph{squared} configurations. These defects each carry a topological charge of $+1/2$. For $m=1$, two reside at the poles and two on the equator ($\vartheta = \pi/2$); for $m=2$, all four lie on the equator. ESM~\ref{sec:rigidtransformation} demonstrates that modes $m=1$ and $m=2$ map onto each other via a rigid transformation. 

We set the imaginary parts of $C_{21}$ and $C_{22}$ to zero without loss of generality:
\begin{equation}
\mathrm{Im}[C_{2i}] = 0, \quad C_i = \mathrm{Re}[C_{2i}], \quad i \in \{1,2\}.
\end{equation}
Any configuration with $\mathrm{Im}[C_{2i}] \neq 0$ represents merely a rigid rotation about the $z$-axis, as the modulus $|C_{2i}|$ coincides with $|C_{i}|$.

For both modes $i \in \{1,2\}$, the cubic energy contribution vanishes ($W^{(3)}[C_i] = 0$), dictating the scaling exponent $\lambda=2$ in~\eqref{eq:a_formula}. The expansion thus reduces to the quartic order:
\begin{subequations}
\label{eq:W3W4C1}
\begin{gather}
W^{(4)}[C_i] = - \frac{2 C_i^2 (\bar \gamma - 5) k \rho_0}{1 + \bar \gamma} 
+ \frac{5 C_i^4 k \rho_0}{14 (1 + \bar \gamma)^4 \pi} \left[ \bar{c} (1 + \bar \gamma)^4 + 54 \big(119 + 28 \bar \gamma + \bar \gamma^2 \big) \right]. \label{eq:W4Ci}
\end{gather}
\end{subequations}
The determination of local minima therefore relies entirely upon the quartic term. Because the resulting amplitudes are strictly real and the prefactor of the quartic term is positive, the bifurcation is supercritical, signifying a continuous, second-order phase transition. The stationary points of $\mathcal{F}$ with respect to $C_i$ yield the isotropic shell:
\be
C_i = 0,
\ee
and the bifurcated shells with the onset of nematic order:
\begin{equation}
\label{eq:c1_anisotropy}
C_{i,\pm}^{\bar \gamma} = \pm \sqrt{\frac{14\pi}{5}} \frac{(1+\bar \gamma)\sqrt{\bar \gamma^2-4\bar \gamma-5}}{\sqrt{\bar{c}(1+\bar \gamma)^4 + 54\,(119 + 28\bar \gamma + \bar \gamma^2)}}.
\end{equation}

Figure~\ref{fig:contourplots_c12} provides three-dimensional renderings of the non-axisymmetric configurations for $C_{i,-}^{\bar \gamma}$ (evaluated at $\bar \gamma^{\mathrm{max}}$). The colour map denotes the magnitude of $q_{1,2i}$, whilst the orange and grey curves delineate the nematic director integral lines for $i=1$ and $i=2$, respectively.  This figure also shows cross-sections along the planes containing the four defects (marked with red circles). Comparing these to the reference sphere (dashed black line) reveals that no deformation occurs exactly at the \emph{half-charged} defects. 

\begin{figure}[t]
   \centering
   \includegraphics[width=0.3\linewidth]{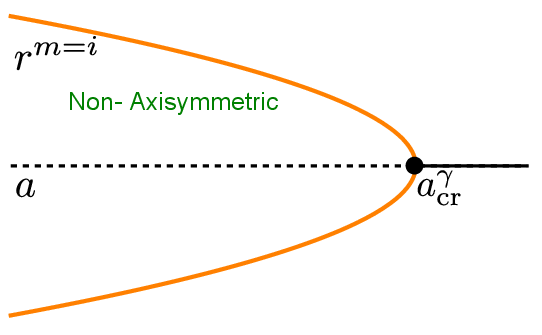}
   \caption{Overview of the second-order bifurcation scenario. For $a \in (0, \astar)$, the functional $\mathcal{F}$ attains a single minimum in $r^{m=i}$ from~\eqref{eq:rm12_def} at $r^{m=i} \equiv r_0$, corresponding to the isotropic branch, which is globally stable. A fold at $a = a_{\mathrm{cr}}^{\bar\gamma}$ gives rise to two additional non-trivial equilibrium solutions. 
   }
    \label{fig:r_epsilon1}
\end{figure}

Analogous to the axisymmetric case, Figure~\ref{fig:r_epsilon1} illustrates the dependence of the equilibrium solutions $r^{m=i}$ \eqref{eq:rm12_def} upon $a$. The eigenfunctions $r_{1,2i}$ are those associated with $C_{i,\pm}^{\bar \gamma}$. The scaling parameter $\varepsilon$ follows \eqref{eq:a_formula} for $\lambda = 2$:
\begin{equation}
\label{eq:varepsilon_lambda2}
\varepsilon = \sqrt{\frac{a - a_{\mathrm{cr}}^{\bar \gamma}}{a_{\mathrm{cr}}^{\bar \gamma}}},
\end{equation}
where $a_{\mathrm{cr}}^{\bar \gamma}$ is given by \eqref{eq:aval2scaled}. 

When the radial functions $r^{m=i}$ are evaluated at the coordinates where their modulus (and thus the displacement) attains its maximum, both solutions yield identical values. Consequently, Figure~\ref{fig:r_epsilon1} plots only one scenario.

In contrast to the axisymmetric case, the critical value $a = a_{\mathrm{cr}}^{\bar \gamma} < \astar$ heralds a continuous, second-order bifurcation. For $0 \ge a \ge a_{\mathrm{cr}}^{\bar \gamma}$, the sole permissible equilibrium is the trivial spherical state $C_i = 0$ ($r^{m=i} \equiv r_0$), depicted in Figure~\ref{fig:figFcomparison} as a horizontal line. At $a = a_{\mathrm{cr}}^{\bar \gamma}$, two additional equilibrium solutions bifurcate from $r_0$, exhibiting perfect symmetry about the horizontal axis ($C_{i,-}^{\bar \gamma} = - C_{i,+}^{\bar \gamma}$).

Evaluating the free-energy $\mathcal{F}$ at $C_{i,\pm}^{\bar \gamma}$ using \eqref{eq:varepsilon_lambda2} yields:
\begin{equation}
\label{eq:energy_second_order}
\mathcal{F}[C_{1,\pm}^{\bar \gamma}] = \mathcal{F}[C_{2,\pm}^{\bar \gamma}] = - \frac{14 \, (-5 - 4 \bar \gamma + \bar \gamma^2)^2 k \pi \rho_0}{5 \left( \bar{c} (1 + \bar \gamma)^4 + 54 (119 + 28 \bar \gamma + \bar \gamma^2) \right)} \left( \frac{a - a_{\mathrm{cr}}^{\bar \gamma}}{a_{\mathrm{cr}}^{\bar \gamma}} \right)^2 < \mathcal{F}[0] = 0,
\end{equation}
which holds whenever the non-trivial solutions exist ($a < a_{\mathrm{cr}}^{\bar \gamma}$). This demonstrates that the trivial solution is rendered metastable at $a = a_{\mathrm{cr}}^{\bar \gamma}$, prompting a \emph{smooth} (second-order) transition towards a non-trivial configuration of lower free energy.

Finally, in the limit $\bar \gamma \to +\infty$, the functional $\mathcal{F}[C_{i,\pm}^{\bar \gamma}]$ reduces to:
\begin{equation}
\label{eq:free_energyinfinityCi}
\mathcal{F}[C_{i,\pm}^{\bar \gamma}] \overset{\bar \gamma\to+\infty}{\longrightarrow} -\frac{14 k \pi \rho_0}{5\bar{c}} \left(\frac{a - a_{\mathrm{cr}}^\infty}{a_{\mathrm{cr}}^\infty}\right)^2, \qquad a_{\mathrm{cr}}^\infty = -\frac{2k}{r_0^2},
\end{equation}
coinciding with the undeformable shell model~\cite{Napoli:2021}. Taking the limit $\bar \gamma \to +\infty$ in~\eqref{eq:c1_anisotropy} similarly recovers the amplitudes for an undeformable shell:
\begin{equation}
C_{i,\pm}^{\infty} = \pm \sqrt{\frac{14 \pi}{5\bar{c}}}.
\end{equation}

\subsection{Energy landscapes of the bifurcated modes} \label{sec:comparison_modes}
We now compare the dimensionless energies of all derived branches to construct a comprehensive bifurcation diagram. Recall that for $a > \astar$, solely the ground-state solution exists. For $a < \astar$, this trivial state coexists with both the prolate and oblate axisymmetric solutions. Finally, for $a < a_{\mathrm{cr}}^{\bar \gamma}$, these are joined by the `squared' solutions.

\begin{figure}[t]
    \centering
    \includegraphics[width=0.4\linewidth]{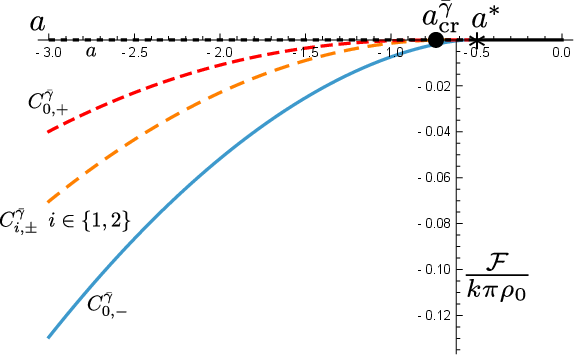}
    \qquad
    \includegraphics[width=0.4\linewidth]{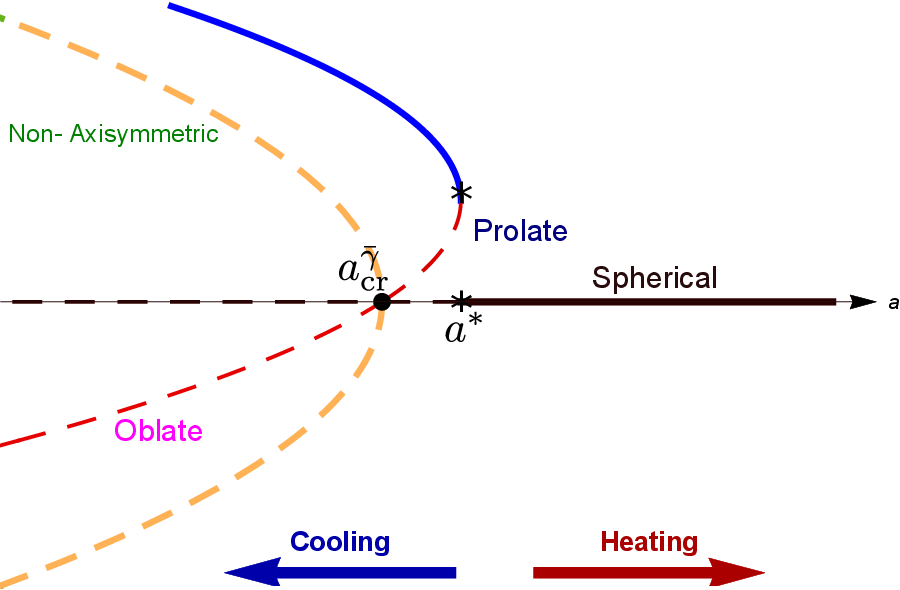}
    \caption{Dimensionless energies $\mathcal{F}/(k\pi \rho_0)$ of the bifurcated configurations reported in Table~\ref{tab:modes_constrained}. The blue line corresponds to the prolate solution associated with $C_{0,-}^{\bar \gamma}$, while the red line corresponds to the solution associated with $C_{0,+}^{\bar \gamma}$, which starts as prolate and subsequently becomes oblate. The orange line represents the squared configurations associated with the modes $m=1$ and $m=2$, which emerge at $a = a_{\mathrm{cr}}^{\bar \gamma}$ and share the same energy. Here, $\bar \gamma = 8$, with $\astar = -0.5$ and $a_{\mathrm{cr}}^{\bar \gamma=8} = -0.67$. The associated complete bifurcation diagram is shown in the right panel.}
    \label{fig:figFcomparison}
\end{figure}

Figure~\ref{fig:figFcomparison} (left panel) illustrates the energy of the oblate configurations (red) and the prolate ones (blue). The energy of the squared solutions---identical for both $m=1$ and $m=2$---lies strictly between the energies of the two axisymmetric states.

We can thus plot the complete bifurcation diagram (Figure~\ref{fig:figFcomparison}, right panel), where solid curves denote the global minimum. During an idealised cooling process, the ground state instability may manifest as early as $a = \astar$, well before the critical value $a_{\mathrm{cr}}^{\bar \gamma}$ predicted by linear analysis. This drives an abrupt transition from the isotropic shape to the prolate configuration (director aligned along meridians).

The remaining non-trivial branches (dashed lines) correspond to local minima and are, in principle, experimentally observable. When multiple local minima coexist, the system's dynamics---not considered within the present static framework---play a decisive role via initial conditions.

Our calculations reveal that the transition's discontinuity is more pronounced for softer shells (lower $\bar \gamma$). In the limit of infinite shell stiffness, even the axisymmetric branch emerges via a supercritical bifurcation, and all identified bifurcated solutions exhibit identical energy levels.

\vspace{-0.481cm}

\subsection{Bifurcation-induced mass redistribution}
\begin{figure}[t]
    \centering
    \begin{subfigure}[c]{0.1\linewidth}
        \centering
        \includegraphics[width=0.4\linewidth]{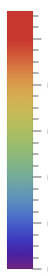}
    \end{subfigure}
        \begin{subfigure}[c]{0.21\linewidth}
        \centering
        \includegraphics[width=0.73\linewidth]{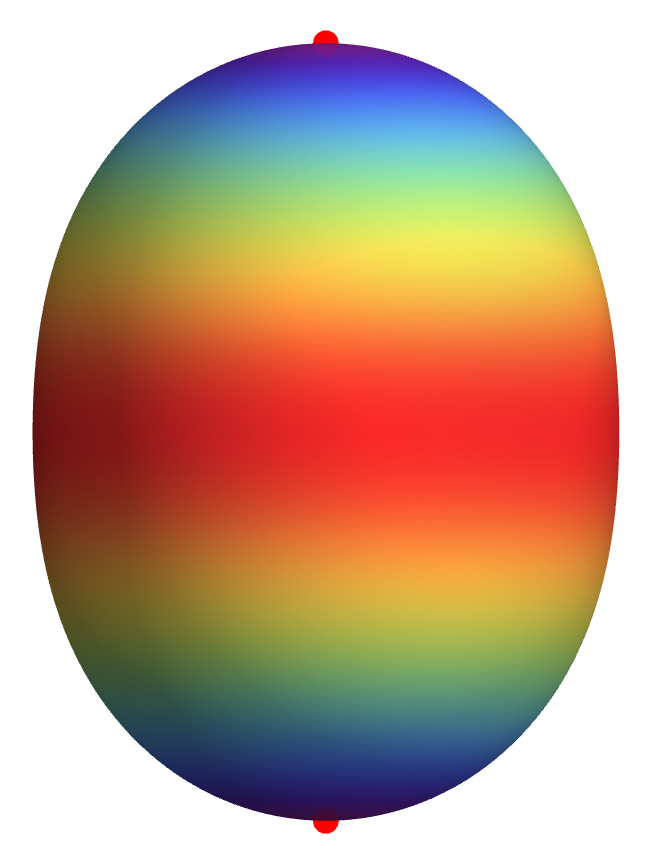}
        \caption{$m=0$ and $C_{0,-}^{\bar \gamma}$.} 
    \end{subfigure}
    \begin{subfigure}[c]{0.21\linewidth}
        \centering
        \includegraphics[width=\linewidth]{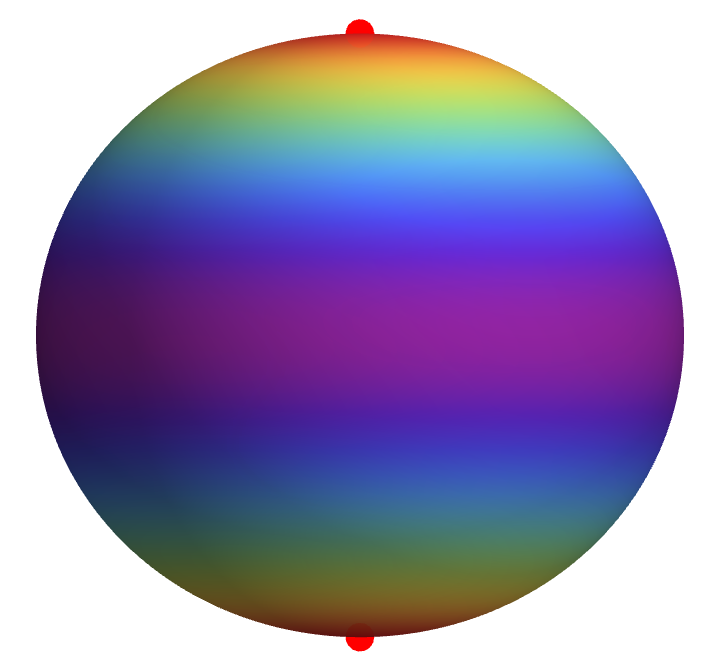}
        \caption{$m=0$ and $C_{0,+}^{\bar \gamma}$.} 
    \end{subfigure}
    \begin{subfigure}[c]{0.21\linewidth}
        \centering
        \includegraphics[width=\linewidth]{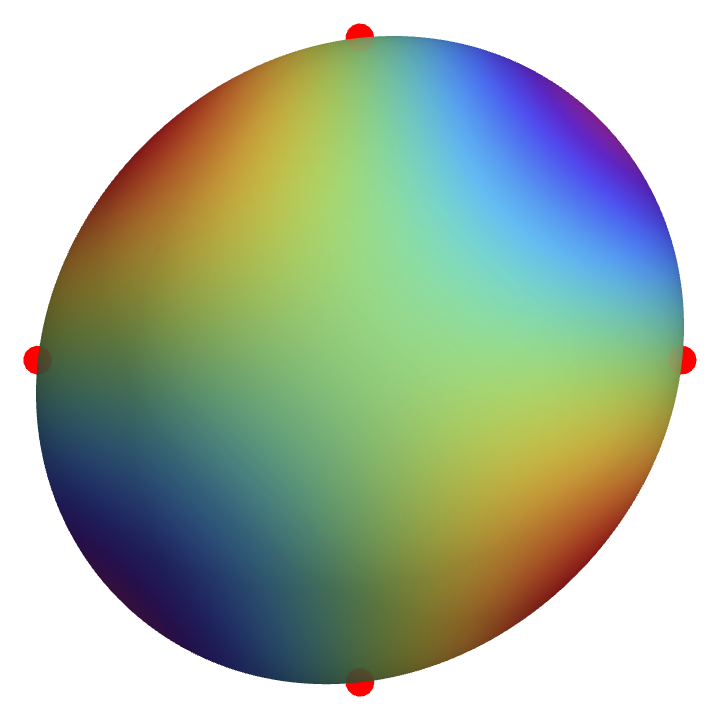}
        \caption{$m=1$ and $C_{1,-}^{\bar \gamma}$.} 
    \end{subfigure}
        \begin{subfigure}[c]{0.21\linewidth}
        \centering
        \includegraphics[width=\linewidth]{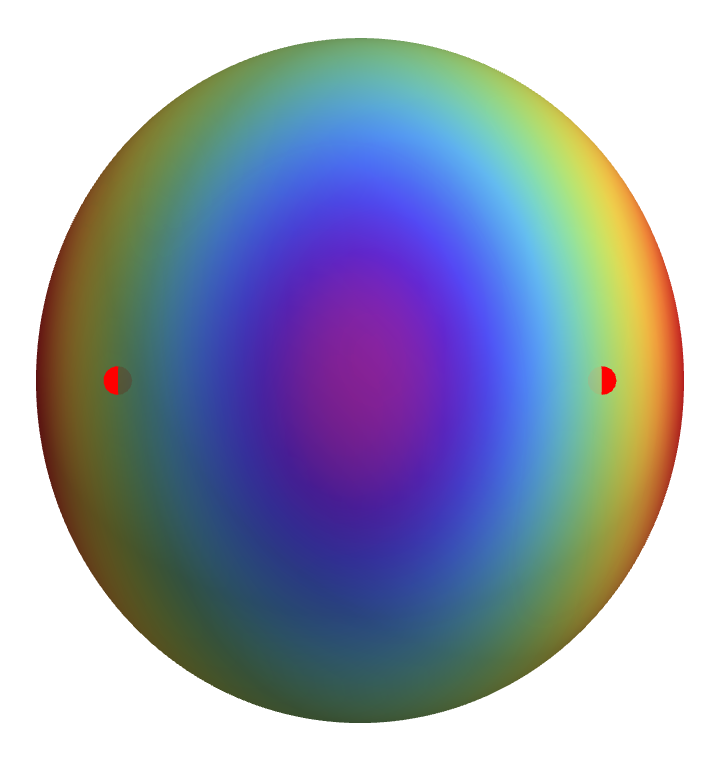}
        \caption{$m=2$ and $C_{2,-}^{\bar \gamma}$.} 
    \end{subfigure}
    \caption{Surface color maps of the local mass density $\rho$ for the bifurcated configurations in Table~\ref{tab:modes_constrained}. Green indicates $\rho = \rho_0$ (no deformation or mass redistribution), violet indicates thinning, and red indicates accumulation. The color legend is shown in the leftmost panel.}
    \label{fig:contourmap_rhof_tot}
\end{figure}

Finally, the surface colour maps in Figure~\ref{fig:contourmap_rhof_tot} illustrate the local surface mass density $\rho$ \eqref{eq:rhoformula} for the bifurcated configurations. Because $\rho$ represents a mass density per unit area, it can be physically interpreted as a change in shell thickness within the three-dimensional continuum: a local increase signifies shell thickening, whilst a decrease corresponds to thinning.

Green regions indicate $\rho = \rho_0$ (no local mass redistribution or shape deformation relative to the undeformed configuration). Darker (violet) regions indicate local mass depletion, coinciding with the most pronounced shape deformation, whilst red regions denote mass accumulation.

Specifically, the prolate configuration exhibits reduced mass density near the poles (along the elongation axis) and mass accumulation near the equator. For the oblate configuration, the converse behaviour occurs: mass density is enhanced near the poles and depleted in the equatorial region, where the shape bulges.

Furthermore, whilst integer defects ($+1$) induce significant local variations in surface mass density, the half-integer defects ($+1/2$) for modes $m=1$ and $m=2$ do not. Indeed, no shape deformation or mass variation occurs at the exact locations of these $+1/2$ defects.

\section{Conclusions}
\label{sec:conclusions}
In this paper, we have formulated and analysed a coupled mechano-nematic model to investigate the symmetry-breaking phase transitions of a deformable spherical shell. Motivated by the early stages of \emph{Hydra} regeneration, wherein a folded, disordered spheroid of tissue spontaneously develops an ordered network of actin fibres, we modelled the onset of nematic order as a temperature-induced bifurcation from an isotropic ground state.

Our weakly non-linear asymptotic analysis reveals a rich physical landscape governed by the delicate interplay between topological constraints, surface elasticity, and nematic ordering. 

The principal findings can be synthesised as follows:
\begin{itemize}
\item[(a)] \emph{A Novel Variational Framework:} The foundation of our study rests upon an original and mathematically rigorous continuum model, derived via a variational energy formulation. By seamlessly integrating the classical capillary energy of a classic bubble soap---governed by surface tension and an internal volume constraint---with the two-dimensional Landau--de Gennes nematic distortion energy, we have established a robust theoretical framework capable of capturing the fully coupled mechano-nematic equilibrium of closed surfaces.
     
\item[(b)] \emph{The Role of Shell Deformability:} The mechanical stiffness of the shell, parametrised by the dimensionless constant $\bar{\gamma}$, fundamentally dictates the nature of the phase transition. For soft, deformable shells, the isotropic-to-nematic transition into an axisymmetric state (featuring two +1 defects) is a discontinuous, transcritical bifurcation. The system abruptly transitions to a state of lower energy prior to reaching the linear critical threshold. However, as the shell approaches the rigid limit ($\bar{\gamma} \to \infty$), the bifurcation gracefully transitions to a continuous, supercritical regime.
    
\item[(c)] \emph{Energy Landscapes and Symmetry Breaking:} We identified distinct equilibrium morphologies corresponding to different azimuthal modes. The global energy minimum is attained by the prolate axisymmetric configuration ($m=0$), wherein the nematic director aligns along the meridians, defining a clear pole-to-pole axis. This demonstrates that the coupling of nematic elasticity and surface deformation provides a purely mechanical pathway for establishing a body axis (head-foot polarity), preceding and potentially guiding subsequent biochemical morphogen gradients. It is also pertinent to note that the fundamental modes analysed herein may be combined to generate more intricate patterns. For instance, a superposition of the two non-axisymmetric modes ($m=1$ and $m=2$)---which, despite representing identical physical configurations when considered in isolation, span a degenerate eigenspace---can yield textures wherein the four +1/2 topological defects are situated at the vertices of a rectangle, rather than a square. This implies that the morphological landscape of the bifurcated shell is further enriched by such symmetry-breaking combinations.

\item[(d)] \emph{Defect-Mass Coupling:} Our model yields novel insights into the interaction between topological charge and surface mass density (which corresponds to shell thickness in the three-dimensional continuum limit). We observed that +1 defects induce significant local variations in mass density, resulting in pronounced thinning or thickening at the poles. By contrast, the half-integer +1/2 defects associated with the non-axisymmetric ``squared'' modes ($m=1, 2$) do not perturb the local mass density in their immediate vicinity, leaving the local shape undeformed relative to the reference sphere.
\end{itemize}

From a biological perspective, these results underscore the capability of topological defects to act as primary organising centres for morphogenesis. The spontaneous elongation of the vesicle and the localisation of stress and mass variations at integer defects mirror the physical phenomena observed during tentacle and axis formation in living organisms.

From a theoretical and computational standpoint, the analytical solutions derived in the present study offer a rigorous benchmark for the validation of numerical codes, particularly in the immediate vicinity of the critical bifurcation threshold. Furthermore, subsequent investigations could fruitfully explore higher-order eigenvalues within the spectrum, which may yield novel classes of solutions featuring negative topological defects. A notable biological analogue is found in mature \emph{Hydra}, where defects of charge -1/2 are observed at the base of emerging protrusions. 

While the present work relies on a static, thermotropic analogy to model the isotropic-nematic transition, it establishes a robust equilibrium framework. Future extensions of this model will incorporate active stresses and dynamic fluid flows, which are essential for capturing the continuous energy consumption and out-of-equilibrium dynamics inherent to living tissues. Integrating this direct coupling between the mechanically induced shape changes and the biochemical expression of morphogens remains a compelling avenue for fully unravelling the mechanics of \emph{Hydra} regeneration.

\subsection*{Acknowledgements}
G.N. and S.P. are member of the Italian Gruppo Nazionale per la
Fisica Matematica (GNFM), which is part of INdAM, the Italian National Institute for Advanced Mathematics. S.P. gratefully
acknowledge partial financial support provided by the INdAM - GNFM Project, CUP code [E5324001950001].

\bibliographystyle{plain} 
\bibliography{ReferencesNB}

\clearpage 
\beginsupplement 

\begin{center}
    \Large \textbf{Electronic Supplementary Material\\
    \vspace{0.5cm}
    Nematic Bubbles}
    
    \vspace{0.5cm}
    
    G. Napoli$^{1}$ and  S. Paparini $^{2}$
    
\vspace{0.2cm}
{\small $^{1}$ Dipartimento di Matematica e Applicazioni "Renato Caccioppoli", Università degli Studi di Napoli "Federico II", Napoli, Italy \\
$^{2}$ Dipartimento di Matematica “Tullio Levi-Civita”, Università degli Studi di Padova,
Padova, Italy}
\end{center}
\vspace{1cm}

\vspace{-1.2cm}

\section{Useful Computations}\label{sec:computations}
Let $(\ev_r, \ev_\vartheta, \ev_\varphi)$ denote the standard spherical frame. We recall that we are considering surfaces $\mathcal{S}$ described by points $\pv = r(\vartheta,\varphi)\ev_r$, as in \eqref{eq:pointp}. Consequently, the explicit expressions of the vectors $\ev_1, \, \ev_2, \, \bnu$ and $\ev_1^\perp$ of the local bases on $\mathcal{S}$ defined in Section~\ref{sec:linearization} are expressed in terms of $r$ as:
\begin{subequations}
\label{eq:evevperpnormal}
\begin{gather}
\ev_1 = \frac{\pv_{,\vartheta}}{|\pv_{,\vartheta}|} = \frac{r_{,\vartheta}\ev_r + r\ev_\vartheta}{\sqrtone} , \\
\ev_2 = \frac{\pv_{,\varphi}}{|\pv_{,\varphi}|} = \frac{r_{,\varphi}\ev_r + r\sin\vartheta\ev_\varphi}{\sqrttwo} , \\
\bnu = \frac{\ev_1 \times \ev_2}{|\ev_1 \times \ev_2|} = \frac{r\sin\vartheta\ev_r - r_{,\vartheta}\sin\vartheta\ev_\vartheta - r_{,\varphi}\ev_\varphi}{\sqrtthree} , \\
\ev_1^\perp = \bnu \times \ev_1 = \frac{r r_{,\varphi}\ev_r - r_{,\varphi}r_{,\vartheta}\ev_\vartheta + \sin\vartheta\one\ev_\varphi}{\sqrtone\sqrtthree} .
\end{gather}
\end{subequations}

Let us formally introduce the operators $\grads$ and $\dvs$, denoting,
respectively, the surface gradient and the surface divergence.

Let $\psi \colon \mathcal{S} \to \mathbb{R}$ be a scalar field of class $\mathcal{C}^1$
defined on $\mathcal{S}$. The surface gradient of $\psi$ at a point
$\pv \in \mathcal{S}$ is a vector $\grads\psi(\pv)$ on the tangent
plane $\mathcal{T}_{\pv}\mathcal{S}$ to $\mathcal{S}$ characterized by the property that,
for any curve $t \to c(t)$ of class $\mathcal{C}^1$
such that $c(t_0) = \pv$, it holds that
\be
\left.\frac{\dd}{\dd t}\psi(c(t))\right|_{t=t_0} = \grads\psi(\pv) \cdot \dot{c}(t_0) .
\ee
For a vector field $\fv \colon \mathcal{S} \to \mathcal{V}$ of class $\mathcal{C}^1$ 
on $\mathcal{S}$, the surface gradient of $\fv$ at a point 
$\pv \in \mathcal{S}$ is defined as the tensor $\grads\fv(\pv)$ 
acting on the tangent plane $\mathcal{T}_{\pv}\mathcal{S}$ such that
\be 
\left.\frac{\dd}{\dd t}\fv(c(t))\right|_{t=t_0} = \bigl( \grads\fv(\pv) \bigr) \dot{c}(t_0) .
\ee

\subsection{Gradient of a scalar field defined on the surface}\label{sec:gradspsi}
Let $\psi$ be a field of class $\mathcal{C}^1$ on $\mathcal{S}$, and $t \to c(t)$ any curve of
class $\mathcal{C}^1$ in $\mathcal{S}$. If we denote by $\dot{\psi}$ the derivative of the scalar-valued function defined by $t \to \psi(c(t))$, then 
\be
\label{eq:dotpsiformula}
\dot{\psi} = \grads\psi \cdot \dot{\pv} ,
\ee
where $\dot{\pv}$ is the derivative of $\pv$. If $\psi$ is expressed in the coordinates $(\vartheta,\varphi)$ on $\mathcal{S}$, then
\be
\label{eq:dotpsi}
\dot{\psi} = \psi_{,\vartheta}\dot{\vartheta} + \psi_{,\varphi}\dot{\varphi} ,
\ee
where $\dot{\vartheta}$ and $\dot{\varphi}$ are the derivatives of the functions which describe
the curve $t \to c(t)$ on $\mathcal{S}$, and \eqref{eq:pointp} implies that
\begin{align}
\label{eq:dotp}
\dot{\pv} &= \sqrtone\,\dot{\vartheta}\ev_1 + \sqrttwo\,\dot{\varphi}\ev_2 \notag \\
&= \sqrtone\,\dot{\vartheta}\ev_1 + \frac{r_{,\vartheta}r_{,\varphi}\ev_1 + r\sqrtthree\ev_1^\perp}{\sqrtone}  \dot{\varphi} ,
\end{align}
where use has been made of
\be
\ev_2 = \frac{\pv_{,\varphi}}{|\pv_{,\varphi}|} = \frac{r_{,\vartheta}r_{,\varphi}\ev_1 + r\sqrtthree\ev_1^\perp}{\sqrtone\sqrttwo} .
\ee
Making use of both \eqref{eq:dotp} and \eqref{eq:dotpsi} in \eqref{eq:dotpsiformula}, since the curve $c$ is
arbitrary we easily arrive at the formula that expresses $\grads\psi$ in the
movable frame:
\be
\label{eq:nablapsi}
\grads\psi = \frac{\psi_{,\vartheta}}{\sqrtone}\ev_1 + \frac{\psi_{,\varphi}\one - \psi_{,\vartheta}r_{,\vartheta}r_{,\varphi}}{r\sqrtone\sqrtthree}\ev_1^\perp .
\ee

\subsection{Gradient of a vector field on a surface}\label{sec:gradsv}

Likewise, if $\fv$ is a vector field of class $\mathcal{C}^1$, we have
\be
\dot{\fv} = \left(\grads\fv\right) \dot{\pv} .
\ee
This formula is to be applied to the unit vectors of the movable
frame $(\ev_1, \ev_1^\perp, \bnu)$. With the aid of
\begin{subequations}
\begin{gather}
\dot{\ev}_1 = \frac{1}{|\pv_{,\vartheta}|}\bm{\mathrm{P}}(\ev_1)\bigl[\pv_{,\vartheta\vartheta}\dot{\vartheta} + \pv_{,\vartheta\varphi}\dot{\varphi}\bigr] , \quad
\dot{\ev}_2 = \frac{1}{|\pv_{,\varphi}|}\bm{\mathrm{P}}(\ev_2)\bigl[\pv_{,\vartheta\varphi}\dot{\vartheta} + \pv_{,\varphi\varphi}\dot{\varphi}\bigr] , \\
\dot{\bnu} = \dot{\ev}_1 \times \ev_2 + \ev_1 \times \dot{\ev}_2 , \quad \dot{\ev}_1^\perp = \dot{\bnu} \times \ev_1 + \bnu \times \dot{\ev}_1 ,
\end{gather}
\end{subequations}
we obtain
\begin{subequations}
\label{eq:gradse1e1perpnormal}
\begin{gather}
\grads\ev_1 = -a_{21}\ev_1^\perp \otimes \ev_1 + a_{22}\ev_1^\perp \otimes \ev_1^\perp + a_{31}\bnu \otimes \ev_1 + a_{32}\bnu \otimes \ev_1^\perp , \\
\grads\ev_2 = a_{21}\ev_1 \otimes \ev_1 - a_{22}\ev_1 \otimes \ev_1^\perp + a_{32}\bnu \otimes \ev_1 + b_{32}\bnu \otimes \ev_1^\perp , \\
\grads\bnu = -a_{31}\ev_1 \otimes \ev_1 - a_{32}\left(\ev_1 \otimes \ev_1^\perp + \ev_1^\perp \otimes \ev_1\right) - b_{32}\ev_1^\perp \otimes \ev_1^\perp ,
\end{gather}
\end{subequations}
with
\begin{align}
a_{21} &= -\frac{r_{,\varphi}\left(r^2+2r_{,\vartheta}^2-rr_{,\vartheta\vartheta}\right)}{\one^{3/2}\sqrtthree} \, , \\[1ex]
a_{22} &= \frac{ \one^2 \sin\vartheta (r\cos\vartheta + r_{,\vartheta}\sin\vartheta) + r\one r_{,\varphi} r_{,\vartheta\varphi} + r_{,\varphi}^2 r_{,\vartheta} (r_{,\vartheta}^2 - r r_{,\vartheta\vartheta}) }{ r\one^{3/2}\three } \, , \\[1ex]
a_{31} &= -\frac{\left(r^2+2r_{,\vartheta}^2-rr_{,\vartheta\vartheta}\right)\sin\vartheta}{\one\sqrtthree} \, , \\[1ex]
a_{32} &= -\frac{\sin\vartheta \bigl[ \one (\cot\vartheta r_{,\varphi} - r_{,\vartheta\varphi}) + r_{,\varphi} r_{,\vartheta} (r + r_{,\vartheta\vartheta}) \bigr]}{r\one^{3/2}\three} \, , \\[1ex]
b_{32} &= \frac{1}{r\one\three^{3/2}} \biggl[ \one^2 \sin\vartheta (r_{,\varphi\varphi} + r_{,\vartheta}\sin\vartheta\cos\vartheta - r\sin^2\vartheta) \notag \\
&\quad - 2\one r_{,\vartheta} r_{,\varphi} r_{,\vartheta\varphi} \sin\vartheta + r_{,\varphi}^2 \bigl( 2\one r_{,\vartheta}\cos\vartheta + \sin\vartheta ( r_{,\vartheta}^2 r_{,\vartheta\vartheta} - 2r^3 - r r_{,\vartheta}^2 ) \bigr) \biggr] .
\end{align}

By \eqref{eq:gradse1e1perpnormal}, the mean curvature $H$ is given by
\be 
\label{eq:HK_appendix}
H = -\frac{1}{2}\dvs\bnu = \frac{1}{2}(a_{31}+b_{32}) .
\ee

\subsection{Surface Divergence}\label{sec:divs}
For any scalar field $\psi$, vectorial fields $\vv$ and $\wv$, and tensorial field $\mathbf{S}$, the following useful identities
can be readily proved:
\begin{subequations}
\label{eq:diff_rules}
\begin{gather}
\dvs(\psi\vv) = \psi\dvs\vv + \grads\psi \cdot \vv , \\
\dvs(\psi\mathbf{S}) = \psi\dvs\mathbf{S} + \mathbf{S}\grads\psi , \\
\dvs(\vv \otimes \wv) = (\dvs\wv)\vv + (\grads\vv)\wv .
\end{gather}
\end{subequations}
Let $v_1$, $v_1^\perp$, and $v_3$ denote the components of $\vv \in \mathcal{V}$ in the movable frame $\ev := (\ev_1, \ev_1^\perp, \bnu)$, so that
\be
\vv = v_1\ev_1 + v_1^\perp\ev_1^\perp + v_3\bnu .
\ee
Thus, by \eqref{eq:gradse1e1perpnormal} and the differentiation rules recalled above, it follows that
\be
\dvs\vv = a_{22}v_1 + a_{21}v_1^\perp - 2H v_3 + \frac{v_{1,\vartheta}}{\sqrtone} + \frac{v_{1,\varphi}^\perp\one - v_{1,\vartheta}^\perp r_{,\vartheta}r_{,\varphi}}{r\sqrtone\sqrtthree} .
\ee

Let now $S_{ij}$ for $i,j=1,2,3$ denote the components of $\mathbf{S} \in \mathcal{L}(\mathcal{V})$ in the basis
$\ev^2 = \{ \ev_i \otimes \ev_j \mid i,j=1,2,3, \text{ and } \ev_i, \, \ev_j \in \ev \}$, so that
\be
\mathbf{S} = \sum_{i,j=1}^3 S_{ij}\, \ev_i \otimes \ev_j .
\ee
By \eqref{eq:gradse1e1perpnormal} and the differentiation rules in \eqref{eq:diff_rules}, it follows that
\begin{align}
\dvs\mathbf{S} &= \biggl[ a_{22}(S_{11}-S_{22}) + a_{21}(S_{12}+S_{21}) - 2H S_{13} - a_{31}S_{31} - a_{32}S_{32} \notag \\
&\qquad + \frac{S_{11,\vartheta}}{\sqrtone} + \frac{S_{12,\varphi}\one - S_{12,\vartheta}r_{,\vartheta}r_{,\varphi}}{r\sqrtone\sqrtthree} \biggr] \ev_1 \notag \\
&\quad + \biggl[ a_{21}(S_{11}+S_{22}) + a_{22}(S_{12}+S_{21}) - 2H S_{23} - a_{32}S_{31} - b_{32}S_{32} \notag \\
&\qquad + \frac{S_{21,\vartheta}}{\sqrtone} + \frac{S_{22,\varphi}\one - S_{22,\vartheta}r_{,\vartheta}r_{,\varphi}}{r\sqrtone\sqrtthree} \biggr] \ev_1^\perp \notag \\
&\quad + \biggl[ a_{31}S_{11} + a_{32}(S_{12}+S_{21}) + b_{32}S_{22} - 2H S_{33} + a_{22}S_{31} + a_{21}S_{32} \notag \\
&\qquad + \frac{S_{31,\vartheta}}{\sqrtone} + \frac{S_{32,\varphi}\one - S_{32,\vartheta}r_{,\vartheta}r_{,\varphi}}{r\sqrtone\sqrtthree} \biggr] \bnu .
\end{align}

\subsection{Jacobian Determinant and Conservation of Mass}\label{sec:mass_conservation}
If $\rho_0$ represents the mass density of the reference sphere, the mass density $\rho(\vartheta,\varphi)$ of the deformed sphere is determined by virtue of the mass-conservation principle, which reads as
\be
\rho(\vartheta,\varphi)\,\dd A = \rho_0\,\dd A_0 ,
\ee
where $\dd A$ and $\dd A_0$ denote the area measures on the deformed and reference spheres, respectively. For a sphere of radius $r_0$, $\dd A_0 = r_0^2 \sin\vartheta \, \dd\vartheta \, \dd\varphi$, while the equation for $\dot{\pv}$ in \eqref{eq:dotp}, expressed in terms of the local basis $(\ev_1, \ev_2)$ on the tangent plane to $\mathcal{S}$, is especially expedient to derive the elementary area $\dd A$ of the shell. More precisely,
\be
\label{eq:ddA}
\dd A = \sqrtone\sqrttwo \, \ev_1 \times \ev_2 \cdot \bnu \, \dd\vartheta \, \dd\varphi = r\sqrtthree \, \dd\vartheta \, \dd\varphi .
\ee
Therefore, $\rho(\vartheta,\varphi)$ is given by
\be
\label{eq:rhoformula}
\rho(\vartheta,\varphi) = \rho_0\frac{r_0^2\sin\vartheta}{r\sqrtthree} .
\ee

\subsection{Expansion in $\varepsilon$}\label{sec:expansionvarepsilon}
Here, we present the first-order expansion in $\varepsilon$ of quantities entering the energy functional and the equilibrium equations:
\begin{gather}
\ev_1 = \ev_\vartheta + \frac{\varepsilon}{r_0}r_{1,\vartheta}\ev_r + \mathcal{O}(\varepsilon^2) , \\
\ev_1^\perp = \ev_\varphi + \frac{\varepsilon}{r_0}\csc\vartheta r_{1,\varphi}\ev_r + \mathcal{O}(\varepsilon^2) , \\
\bnu = \ev_r + \frac{\varepsilon}{r_0} \bigl( -r_{1,\vartheta}\ev_\vartheta - \csc\vartheta r_{1,\varphi}\ev_{\varphi} \bigr) + \mathcal{O}(\varepsilon^2) .
\end{gather}

\begin{gather}
\grads\psi = \gradstar\psi + \varepsilon \biggl[ \bigl( \gradstar\psi \cdot \gradstar r_1 \bigr) \ev_r - \frac{r_1}{r_0}\gradstar\psi \biggr] + \mathcal{O}(\varepsilon^2) , \\
a_{21} = \frac{r_{1,\varphi}}{r_0^2\sin\vartheta}\varepsilon + \mathcal{O}(\varepsilon^2) , \\
a_{22} = \frac{\cot\vartheta}{r_0} + \frac{-\cot\vartheta r_1 + r_{1,\vartheta}}{r_0^2}\varepsilon + \mathcal{O}(\varepsilon^2) , \\
a_{31} = -\frac{1}{r_0} + \frac{r_1 + r_{1,\vartheta\vartheta}}{r_0^2}\varepsilon + \mathcal{O}(\varepsilon^2) , \\
a_{32} = -\frac{-\cot\vartheta r_{1,\varphi} + r_{1,\vartheta\varphi}}{r_0^2\sin\vartheta}\varepsilon + \mathcal{O}(\varepsilon^2) , \\
b_{32} = -\frac{1}{r_0} + \frac{1}{r_0} \biggl( r_1 + \frac{r_{1,\varphi\varphi}}{\sin^2\vartheta} + \cot\vartheta r_{1,\vartheta} \biggr) \varepsilon + \mathcal{O}(\varepsilon^2) .
\end{gather}

Let us set
\be
s_1 = 2r_1 + \cot\vartheta \, r_{1,\vartheta} + r_{1,\vartheta\varphi} + \csc^2\vartheta \, r_{1,\varphi\varphi} ;
\ee
we then obtain from \eqref{eq:HK_appendix}
\be
H = - \frac{1}{r_0} + \varepsilon \frac{s_1}{2 r_0^2} + \mathcal{O}(\varepsilon^2) , \qquad K = \frac{1}{r_0^2} - \varepsilon \frac{s_1}{2 r_0^3} + \mathcal{O}(\varepsilon^2) .
\ee

Finally,
\begin{gather}
\dd A = (r_0^2 + 2\varepsilon r_0 r_1) \sin\vartheta \, \dd\vartheta \, \dd\varphi + \mathcal{O}(\varepsilon^2) , \\
\rho = \rho_0 \biggl( 1 - 2\frac{r_1}{r_0} \biggr) .
\end{gather}
\section{Spin-weighted spherical harmonics} \label{sec:spin_weighted}
To rigorously describe tangential tensor fields on a spherical surface, such as the nematic order tensor $\Qv_s$, it is highly advantageous to employ spin-weighted spherical harmonics. Following the seminal formulation by Goldberg \emph{et al.}~\cite{Goldberg:1967}, a complex-valued function $\eta(\vartheta, \varphi)$ defined on $\mathbb{S}^2$ is said to possess spin-weight $s$ if, under a local rotation of the orthonormal tangent frame by an angle $\psi$, it transforms according to $\eta \to e^{i s \psi} \eta$.

The complex scalar amplitude $\QQ$, representing the symmetric and traceless second-order nematic tensor in the tangent plane, naturally transforms with spin-weight $s = \pm 2$. 

The spin-weighted spherical harmonics ${}_\sigma Y_{l,m}(\vartheta, \varphi)$, defined for integers $l \ge |s|$ and $|m| \le l$, are the regular eigenfunctions of the self-adjoint operator $\bar{\eth}\eth$:
\begin{equation}
\bar{\eth}\eth \left( {}_\sigma Y_{l,m} \right) = -(l-\sigma)(l+\sigma+1) \, {}_\sigma Y_{l,m}.
\end{equation}

For $s=0$, these functions reduce to the standard spherical harmonics $Y_{l,m}$. For $s > 0$, they may be generated directly from the standard harmonics via repeated application of the raising operator:
\begin{equation}
{}_\sigma Y_{l,m} = \sqrt{\frac{(l-\sigma)!}{(l+\sigma)!}} \, \eth^s Y_{l,m}.
\end{equation}
Conversely, for $s < 0$, they are obtained by analogous application of the lowering operator $\bar{\eth}$. Crucially, these functions provide a complete and orthonormal basis for square-integrable functions of spin-weight $s$ on the sphere, satisfying the orthogonality condition:
\begin{equation}
\int_0^{2\pi} \int_0^\pi {}_\sigma Y_{l,m} \; {}_\sigma Y^*_{l',m'}  \sin\vartheta \, \dd\vartheta \dd\varphi = \delta_{ll'} \delta_{mm'},
\end{equation}
where the asterisk denotes complex conjugation and $\delta$ is the Kronecker delta.

Table~\ref{tab:modes} details the explicit forms of the spin-weighted spherical harmonics corresponding to $\sigma=0$ and $\sigma=2$ for $l=2$.
\begin{table}[t]
\centering
\renewcommand{\arraystretch}{2.5} 
\begin{tabular}{@{} c c c @{}}
\hline
\textbf{$m$} & { $Y_{2,m}(\vartheta, \varphi)$} & { ${}_2Y_{2,m}(\vartheta, \varphi)$} \\ \hline
$2$  & $ \frac{1}{4}\sqrt{\frac{15}{2\pi}} \sin^2 \vartheta \, \text{e}^{ 2 i \varphi } $       & $ \frac{1}{8}\sqrt{\frac{5}{\pi}} (1+\cos \vartheta )^2 \text{e}^{ 2 i \varphi } $ \\
$1$  & $ -\frac{1}{2}\sqrt{\frac{15}{2\pi}} \sin \vartheta \cos \vartheta \, \text{e}^{i \varphi} $ & $ \frac{1}{4}\sqrt{\frac{5}{\pi}} \sin \vartheta (1+\cos \vartheta ) \text{e}^{i \varphi} $ \\
$0$  & $ \frac{1}{4}\sqrt{\frac{5}{\pi}} (3\cos^2 \vartheta - 1) $                         & $ \frac{1}{4}\sqrt{\frac{15}{2\pi}} \sin^2 \vartheta $ \\
$-1$ & $ \frac{1}{2}\sqrt{\frac{15}{2\pi}} \sin \vartheta \cos \vartheta \, \text{e}^{-i \varphi} $ & $ \frac{1}{4}\sqrt{\frac{5}{\pi}} \sin \vartheta (1-\cos \vartheta ) \text{e}^{-i \varphi} $ \\
$-2$ & $ \frac{1}{4}\sqrt{\frac{15}{2\pi}} \sin^2 \vartheta \, \text{e}^{- 2 i \varphi} $       & $ \frac{1}{8}\sqrt{\frac{5}{\pi}} (1-\cos \vartheta )^2 \text{e}^{- 2 i \varphi} $ \\ \hline
\end{tabular}
\caption{Spin-weight 0 and 2 spherical harmonics with $l=2$.}
\label{tab:modes}
\end{table}

\section{Modal Analysis}\label{sec:modalanalysis_appendix}

This analysis consists in substituting the expressions obtained for the elementary area $\dd A$ in \eqref{eq:ddA}, the mass density $\rho$ in \eqref{eq:rhoformula}, the outer normal to the shell $\normal$ in \eqref{eq:evevperpnormal}, the director field $\n$ in \eqref{eq:n_def} and the parameter $a$ in \eqref{eq:a_formula} into the free-energy functional \eqref{eq:free_energy}.
After these substitutions, the free-energy $W$ becomes a function of the unknown functions $r$, $q$, and $\alpha$. We then insert the expressions for the perturbed radius $r(\vartheta,\varphi)$ and the order parameter $q(\vartheta,\varphi)$ in terms of $\varepsilon$ given in \eqref{eq:q_r_perturbed}.
Finally, we expand the free energy in a power series in $\varepsilon$ and we start by analyzing the contributions of order $0, \, 1,$ and $2$ in this expansion.

The zeroth-order contribution, $W^{(0)}$, is given by the constant
\begin{equation}
\label{eq:W0}
W^{(0)} = 4\pi \left( \gamma r_0^2 + \frac{\mu}{3} R^3 + k \rho_0 \right),
\end{equation}
and therefore it does not contribute to the minimization procedure.

The first-order contribution, instead, reads as
\be
\label{eq:W1}
W^{(1)}=\int_0^{2\pi}\int_{0}^\pi
\frac{1}{r_0 \sin\vartheta} \left[
(2 \gamma r_0^2 + \mu r_0^3 - 2 k \rho_0)\, r_1
- k \rho_0 r_0^2\Deltas r_1\right]\dd\vartheta\dd\varphi,
\ee
where the differential operator $\Deltastar_s$ is the Laplace-Beltrami operator 
on a spherical surface of radius $r_0$, defined as
\be
\Deltastar_s=\frac{1}{r_0^2}\left( 
\csc^2\vartheta \,\partial^2_{\varphi\varphi} 
+ \cot\vartheta \partial_\vartheta+ \partial^2_{\vartheta\vartheta}\right).
\ee
The Euler Lagrange equations with respect to $r_1$ of  \eqref{eq:W1} gives exactly the Young-Laplace law \eqref{eq:YL}.

For brevity, we do not report the explicit form of the second-order functional $W^{(2)}$. It is sufficient to know that the Euler-Lagrange equations with respect to $q$ and $\alpha$ yield, respectively, equations \eqref{eq:qlin} and \eqref{eq:nlin}. Moreover, the Euler-Lagrange equation with respect to $r_1$ of $W^{(2)}$ gives \eqref{eq:r1EL}. 
This implies that, upon substituting $a_{\rm cr}$ from \eqref{eq:a2def} and $\mu$ as in \eqref{eq:YL}, and by considering from \eqref{eq:eigenfunctions_general}
\begin{equation}
r_1(\vartheta,\varphi) = \sum_{m=0}^{3} r_{1,2m}, \quad
q_1(\vartheta,\varphi)=\sqrt{Q Q^*}, \quad
\alpha(\vartheta,\varphi)= \frac{1}{2} \arctan\left(\frac{\mathrm{Im}[Q]}{\mathrm{Re}[Q]}\right)
\end{equation}
with
\be
Q(\vartheta,\varphi) = \sum_{m=0}^{3} Q_{2m}, 
\ee
and where the eigenfunctions $r_{1,2m}$ and $Q_{2m}$ for $m \in \{0,1,2\}$ are listed in Table ..., both the first and the second-order contributions, $W^{(1)}$ and $W^{(2)}$, vanish.

To sum up, upon performing all these substitutions in the free energy, we obtain the real-valued function $\mathcal{F}$ in \eqref{eq:thirdfourthorder}, up to the additive constant $W^{(0)}$ in \eqref{eq:W0}.

\subsection{Rigid Transformation}\label{sec:rigidtransformation}

We now show that the non-axisymmetric configurations introduced in Sec.~\ref{sec:squared}, corresponding to $m=1$ and $m=2$, are equivalent up to a rigid transformation. 

Without loss of generality, we consider the rigid transformation that maps the point defects of the configuration with $m=1$ and $\mathrm{Im}[Q_{12}]=0$ onto the point defects of the configuration with $m=2$ and $\mathrm{Im}[Q_{22}]=0$. As mentioned in the main text, configurations with $\mathrm{Im}[Q_{1i}]\neq 0$, $i=1,2$, can be obtained by a rigid rotation around the $z$-axis from the corresponding configurations with $\mathrm{Im}[Q_{1i}]=0$, $i=1,2$.

The rigid transformation relevant to our case consists of a rotation by $\vartheta=-\pi/2$ about the $x$-axis, followed by a rotation by $\varphi=\pi/4$ about the resulting $z$-axis; it is represented by the tensor
\begin{equation}
\bold{R} = \frac{1}{\sqrt{2}}\left(\mathbf{e}_x \otimes \mathbf{e}_x - \mathbf{e}_x \otimes \mathbf{e}_z + \mathbf{e}_y \otimes \mathbf{e}_x + \mathbf{e}_y \otimes \mathbf{e}_z \right) - \mathbf{e}_z \otimes \mathbf{e}_y.
\end{equation}

Under this transformation, a generic point $\mathbf{p} = \mathbf{p}(\vartheta,\varphi)$ of the initial configuration is transformed to
\begin{equation}
\label{eq:pprime}
\mathbf{p}^\prime(\vartheta^\prime,\varphi^\prime) = \mathbf{R} \mathbf{p}(\vartheta,\varphi),
\end{equation}
where $(\vartheta^\prime,\varphi^\prime)$ are the coordinates of $\mathbf{p}^\prime$ in the \emph{dragged} local basis 
$(\mathbf{e}^\prime_r,\mathbf{e}^\prime_\varphi,\mathbf{e}^\prime_\vartheta)$, obtained from the original basis 
$(\mathbf{e}_r,\mathbf{e}_\varphi,\mathbf{e}_\vartheta)$ at $\mathbf{p}$ with $\mathbf{e}^\prime_i =  \mathbf{R} \mathbf{e}_i$.
To obtain the change of coordinates described by $(\widehat{\vartheta}(\vartheta',\varphi'), \widehat{\varphi}(\vartheta',\varphi'))$,
it then suffices to solve
$\mathbf{e}_r^\prime = \bold{R} \mathbf{e}_r$, which holds if and only if
\begin{equation}
\begin{cases}
\cos\varphi^{\prime}\sin\vartheta^{\prime}=\dfrac{-\cos\vartheta + \cos\varphi\,\sin\vartheta}{\sqrt{2}}, \\[2mm]
\sin\varphi^{\prime}\sin\vartheta^{\prime}=\dfrac{\cos\vartheta + \cos\varphi\,\sin\vartheta}{\sqrt{2}}, \\[1mm]
\cos\vartheta^{\prime}=-\sin\varphi\,\sin\vartheta.
\end{cases}
\end{equation}
The solutions in $\varphi$ and $\vartheta$ are
\begin{equation}
\label{eq:varphithetaprime}
\begin{cases}
\displaystyle
\hat{\varphi} =
\arctan\Biggl[
\pm\,\frac{(\cos\varphi' + \sin\varphi')\sin\vartheta'}
{\sqrt{3 - \cos(2\vartheta')(-1 + \sin(2\varphi')) + \sin(2\varphi')}},\\[1mm]
\qquad\quad  \pm\,\dfrac{\cos\vartheta'\,
\sqrt{2 - \cos^2\varphi'\sin^2\vartheta'
      - \sin^2\varphi'\sin^2\vartheta'
      + \sin(2\varphi')\sin^2\vartheta'}}
{-2 + \cos^2\varphi'\sin^2\vartheta'
    - 2\cos\varphi'\sin\varphi'\sin^2\vartheta'
    + \sin^2\varphi'\sin^2\vartheta'}
\Biggr] + 2\pi k, \\[2mm]
\displaystyle
\hat{\vartheta} =
\arctan\Biggl[
-(\cos\varphi' - \sin\varphi')\sin\vartheta',\\[1mm]
\qquad\quad \pm\,\sqrt{2 - \cos^2\varphi'\sin^2\vartheta'
      - \sin^2\varphi'\sin^2\vartheta'
      + \sin(2\varphi')\sin^2\vartheta'}
\Biggr] + 2\pi m,
\end{cases}
\qquad k,m\in\mathbb{Z}.
\end{equation}
By expressing $\mathbf{p}$ and $\mathbf{p}^\prime$ in terms of the corresponding local bases as in \eqref{eq:rm12_def}, we note that \eqref{eq:pprime} holds if and only if
\begin{equation}
r_1^{m=1}(\hat{\vartheta}, \hat{\varphi})=r_1^{m=2}\bigl(\vartheta^\prime, \varphi^\prime\bigr).
\end{equation}
By substituting $(\hat{\vartheta},\hat{\varphi})$ from \eqref{eq:varphithetaprime} into $r_1^{m=2}$, one recovers exactly this result.

The transformation of the nematic order tensor requires more careful consideration. The coordinates $\varphi^\prime$ and $\vartheta^\prime$ refer to the basis dragged by the rigid transformation $\mathbf{R}$, which does not coincide with the \emph{natural} basis in which $Q_{22}$ is expressed in terms of the components of $\Qv_s$.
Let $\ev_{\vartheta}^\prime = \mathbf{R} \ev_{\vartheta}$ and $\ev_{\varphi}^\prime = \mathbf{R} \ev_{\varphi}$ be the unit vectors of the dragged basis, and let $\tilde\ev^\prime_{\vartheta}$ and $\tilde\ev^\prime_{\varphi}$ denote the unit vectors of the natural basis. Since both bases are defined at the same point $\pv^\prime$, they differ by a local rotation of an angle $\psi(\vartheta^\prime,\varphi^\prime)$. This implies that the two representations of the transformation of $\Qv$ 
with respect to the two local bases are related by
\begin{equation}
\begin{pmatrix} \tilde{q}_1 & \tilde{q}_2 \\ \tilde{q}_2 & -\tilde{q}_1 \end{pmatrix} 
= 
\begin{pmatrix} \cos\psi & \sin\psi \\ -\sin\psi & \cos\psi \end{pmatrix}
\begin{pmatrix} q^\prime_1 & q^\prime_2 \\ q^\prime_2 & -q^\prime_1 \end{pmatrix}
\begin{pmatrix} \cos\psi & -\sin\psi \\ \sin\psi & \cos\psi \end{pmatrix}.
\end{equation}
Considering the complex representation $\QQ = q_1 + i q_2$, one obtains that $\tilde{Q}_{22}$ must be expressed in terms of $Q_{21}(\vartheta^\prime, \varphi^\prime)$ as
\begin{equation}
\label{eq:Q22tilde}
\tilde{Q}_{22}=\tilde{q}_1 + i \tilde{q}_2= (\cos 2\psi - i \sin 2\psi)({q}_1^\prime + i {q}_2^\prime) = e^{-2i\psi(\vartheta^\prime, \varphi^\prime)} Q_{21}(\vartheta^\prime, \varphi^\prime).
\end{equation}
Consequently, to verify this, we compute the ratio between the two fields expressed in the transported and natural bases, respectively, and we obtain
\begin{equation}
\frac{{{Q}}_{22}(\vartheta^\prime, \varphi^\prime)}{Q_{21}(\hat{\vartheta}, \hat{\varphi})}=-\frac{\left(3 + \cos(2\vartheta^\prime)\right)\sin(2\varphi^\prime) 
+ 2\sin^2(\vartheta^\prime)}
{3 + \cos(2\vartheta^\prime) 
+ 2\sin(2\varphi^\prime)\sin^2(\vartheta^\prime)}-i\frac{4\cos(2\varphi^\prime)\cos(\vartheta^\prime)}
{3 + \cos(2\vartheta^\prime) + 4\cos(\varphi^\prime)\sin(\varphi^\prime)\sin^2(\vartheta^\prime)}.
\end{equation}
We note that this ratio is a pure phase factor, i.e., a complex number of unit modulus. 
This confirms, in view of \eqref{eq:Q22tilde}, that $Q_{21}$ and $Q_{22}$ represent the same physical configuration.

\section{Influence of the parameter $\bar{\gamma}$}
Whilst the main text analysed the bifurcation diagram encompassing both axisymmetric and non-axisymmetric solution branches with respect to the parameter $a$ near the critical value $a_{\mathrm{cr}}$, we now investigate the dependence of the equilibrium solutions upon $\bar{\gamma} > 0$.

\subsubsection{Axisymmetric configurations}

\begin{figure}[t]
	\centering
	\includegraphics[width=0.45\linewidth]{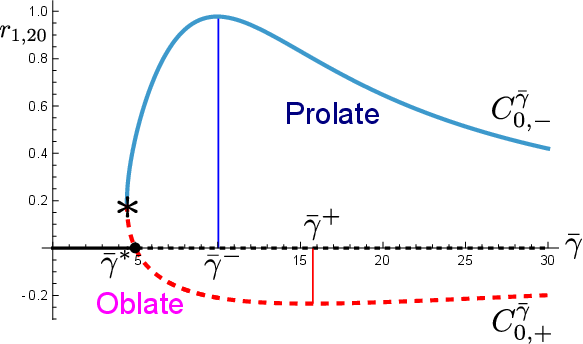}
	\caption{Bifurcation branches of $r_{1,20}$ at the pole $\vartheta = 0$ as a function of $\bar{\gamma}$ for $\varepsilon = 0.1$. The fold at $\gamma^\ast < 5$ gives rise to two additional equilibrium branches: the blue branch $C_{0,-}^{\bar{\gamma}}$ becomes stable, whilst the red branch $C_{0,+}^{\bar{\gamma}}$ remains metastable. The horizontal line denotes the trivial isotropic solution, which becomes metastable at $\gamma^\ast$.}
    \label{fig:figr1c0vsgamma0}
\end{figure}

Fixing $\varepsilon \ll 1$ to approximate the regime near the isotropic–nematic phase transition, Figure~\ref{fig:figr1c0vsgamma0} illustrates the dependence of the equilibrium solution $r^{m=0}$ \eqref{eq:rm0_def}, evaluated at the pole $\vartheta=0$, upon the parameter $\bar\gamma>0$. Specifically, we examine the radial profiles $r_{1,20}$ associated with the equilibrium amplitudes $C_0$ defined in \eqref{eq:c0_isotropy} and \eqref{eq:c0_anisotropy}.  

The horizontal line $r=r_0$ denotes the trivial spherical solution, valid for all $\bar\gamma>0$. The transcritical nature of the bifurcation from the isotropic state to the nematic phase is reflected in the non-symmetric dependence of the equilibrium branches of $r_{1,20}$ upon $\bar{\gamma}$. Furthermore, we identify a lower bound $\gamma^\ast < 5$, dependent upon $a$ and $c$, below which bifurcated solutions cease to exist. 

At this fold (or limit point) $\gamma^\ast$, two additional equilibrium solutions emerge: the blue branch corresponding to $C_{0,-}^{\bar{\gamma}}$ and the red branch corresponding to $C_{0,+}^{\bar{\gamma}}$. As established via the evaluation of the functional $\mathcal{F}$ in the main text, the blue branch immediately becomes the global minimiser at $\gamma^\ast$, thereby supplanting the isotropic branch. Conversely, the red branch remains metastable for all $\bar{\gamma} > \gamma^\ast$, despite possessing a lower energy than the isotropic state.

For all equilibrium branches, as $\bar{\gamma} \to \infty$, $r_1$ tends to zero and the deformed shape $r^{m=0}$ \eqref{eq:rm0_def} approaches a spherical shell, consistent with the undeformable limit. Moreover, Figure~\ref{fig:figr1c0vsgamma0} reveals that along the $C_{0,-}^{\bar{\gamma}}$ branch, $r_{1,20}(0)$ attains a maximum at $\bar{\gamma}^-$, corresponding to the \emph{largest prolate displacement} relative to the reference sphere. Similarly, along the $C_{0,+}^{\bar{\gamma}}$ branch, $r_{1,20}(0)$ exhibits a minimum at $\bar{\gamma}^+$, corresponding to the \emph{largest oblate displacement}.

\subsection{Squared configurations}
\begin{figure}[htbp]
    \centering
    \begin{subfigure}{0.6\textwidth}
        \centering
        \includegraphics[width=0.6\textwidth]{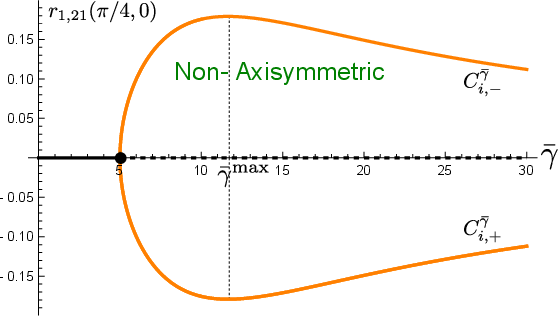}
		\caption{Bifurcated branches $r_{1,21}$ at $(\vartheta,\varphi)=(\pi/4,0)$ as functions of $\bar \gamma$.} 
		\label{fig:figr1c12vsgamma0}
     \end{subfigure}
   \begin{subfigure}{0.19\textwidth}
        \centering
        \includegraphics[width=\textwidth]{r1c1crosssections.eps}
		\caption{$m=1.$} 
		\label{fig:m1crossdefects}
        \end{subfigure}
        \begin{subfigure}{0.19\textwidth}
        \centering
         \includegraphics[width=\textwidth]{r1c2crosssections.eps}
		\caption{$m=2$.} 
		\label{fig:m2crossdefects}
    \end{subfigure}
    \caption{ Black dashed lines in Figs.~\ref{fig:m1crossdefects} and~\ref{fig:m2crossdefects} indicate the cross section of the reference sphere}
	\label{fig:r1energyc0}
\end{figure}

Figure~\ref{fig:figr1c12vsgamma0} illustrates the equilibrium eigenfunctions $r_{1,21}$ \eqref{eq:rm12_def} associated with $C_{1,\pm}^{\bar{\gamma}}$, evaluated at $(\vartheta, \varphi) = (\pi/4, 0)$ and plotted as functions of $\bar{\gamma}$. The second-order nature of the bifurcation is similarly reflected in this dependence upon $\bar{\gamma} > 0$. 

A fold manifests at $\bar{\gamma} = 5$, giving rise to two additional equilibrium solutions emerging from the isotropic state $r_{1,21} \equiv 0$: the upper branch corresponding to $C_{1,-}^{\bar{\gamma}}$ and the lower branch to $C_{1,+}^{\bar{\gamma}}$. These branches exhibit perfect odd symmetry about the horizontal axis and, as established in the main text, possess a lower energy than the trivial solution. Consequently, the isotropic state is rendered metastable at $\bar{\gamma} = 5$.
 
Consistent with the non-deformable limit ($\bar{\gamma} \to +\infty$), the equilibrium solutions $r_1^{m=1}$ converge to $0$. However, there exists a specific value of $\bar{\gamma}$, denoted $\bar{\gamma}^{\rm max}$, at which the absolute displacement from the spherical reference state is maximised.

Figures~\ref{fig:m1crossdefects} and~\ref{fig:m2crossdefects} present the cross-sections of the equilibrium deformed shapes associated with $C_{1,-}^{\bar \gamma}$ and $C_{2,-}^{\bar \gamma}$ at $\bar \gamma=\bar \gamma^{\rm max}$, taken along the planes containing the four defects located at the vertices of a square (marked by red circles). Comparing these profiles with the reference sphere cross-section (dashed black line) reveals that no displacement occurs at the exact locations of these \emph{half-charged} defects.

\end{document}